\documentclass[journal]{IEEEtran}
\usepackage{amsmath,amsfonts}
\usepackage{algorithmic}
\usepackage{array}
\usepackage[caption=false,font=normalsize,labelfont=sf,textfont=sf]{subfig}
\usepackage{textcomp}
\usepackage{stfloats}
\usepackage{url}
\usepackage{verbatim}
\usepackage{graphicx}
\usepackage{enumitem}
\usepackage{svg}
\DeclareMathSizes{10}{10}{4}{2}

\def\BibTeX{{\rm B\kern-.05em{\sc i\kern-.025em b}\kern-.08em
    T\kern-.1667em\lower.7ex\hbox{E}\kern-.125emX}}
\usepackage{balance}
\begin{document}
\title{Reliable Wireless Networking via Soft-Source Information Combining}
\author{Lihao~Zhang and~Soung~Chang~Liew,~\IEEEmembership{Fellow,~IEEE}
\thanks{This work was supported in part by xxx. \textit{(Corresponding author: Soung Chang Liew.)}}
\thanks{Lihao Zhang and Soung Chang Liew are with Department of Information Engineering, The Chinese University of Hong Kong, Hong Kong SAR, China (email: zl018@ie.cuhk.edu.hk; soung@ie.cuhk.edu.hk).}}%
\markboth{Journal of \LaTeX\ Class Files,~Vol.~xx, No.~x, April~2022}%
{Reliable Wireless Networking via Soft-Source Information Combining}

\maketitle

\begin{abstract}
This paper puts forth a multi-stream networking paradigm, referred to as soft-source-information-combining (SSIC), to support wireless Internet of Things (IoT) applications with ultra-reliability requirements. For SSIC networking, an SSIC dispatcher at the source dispatches duplicates of packets over multiple streams, which may be established over different physical wireless networks. If a packet on a stream cannot be decoded due to wireless interference or noise, the decoder makes available the packet’s soft information. An aggregator then combines the soft information of the duplicates to boost reliability. Of importance are two challenges: i) how to descramble the scrambled soft information from different streams to enable correct SSIC; ii) the construct of an SSIC dispatching and aggregation framework compatible with commercial network interface cards (NICs) and TCP/IP networks. To address the challenges, we put forth: i) a soft descrambling (SD) method to minimize the bit-error rate (BER) and packet-error rate (PER) at the SSIC’s output; ii) an SSIC networking architecture readily deployable over today’s TCP/IP networks without specialized NICs. For concept proving and experimentation, we realized an SSIC system over two Wi-Fi’s physical paths in such a way that all legacy TCP/IP applications can enjoy the reliability brought forth by SSIC without modification. Experiments over our testbed corroborate the effectiveness of SSIC in lowering the packet delivery failure rate and the possibility of SSIC in providing 99.99\% reliable packet delivery for short-range communication.
\end{abstract}


\section{Introduction} \label{sec:introduction}
\IEEEPARstart{W}{ireless} networks are error-prone and may fail to deliver packets to their destinations due to environmental interference and noise \cite{aguayo2004link,rodrig2005measurement}. Improving the reliability of wireless networks is essential for the support of mobile industrial, medical, audiovisual and IoT applications with stringent quality-of-service requirements \cite{gokalgandhi2021reliable, wanasinghe2020internet,park2020wireless,savazzi2014wireless,ghoumid2021protocol}.

Many industrial IoT (IIoT) applications call for highly reliable low-latency communications \cite{ma2019high}. A way to ensure reliability is via ``time diversity'': if a message fails to be delivered, just retransmit it. For a wireless network, however, the poor channel condition may persist for a while. For example, the wireless network may be busy due to access by many devices, the wireless path to the device may be blocked, or the device moves to a location far away from the access point (AP). Thus, time diversity may come with the cost of unacceptably high latency. An alternative is to leverage ``network diversity'' or ``path-diversity'', whereby the same message is sent over multiple wireles networks.

``Parallel Redundancy Protocol (PRP)'' \cite{international2012industrial} is an IEC 62439-3 industry standard originally targeted for Ethernet. PRP realizes network redundancy by dispatching duplicate packets over two independent networks. The same concept can be applied to wireless networks. For example, \cite{cena2016experimental} investigated using PRP to enhance the reliability of Wi-Fi networks, whereby each Wi-Fi device is armed with two radio transceivers operating over different channels and bands. 
 
The upcoming IEEE 802.11be ETH (Wi-Fi 7) standard aims to improve Wi-Fi reliability via multi-link operation \cite{khorov2020current}. While the detailed formal standard will only be released in 2024, proponents in \cite{80211be_Task_cb} have already begun to discuss the possibility of integrating IEEE 802.1CB, a standard akin to PRP, into the IEEE 802.11 standard. IEEE 802.1CB is also called frame-replication-and-elimination scheme for reliability (FRER). With IEEE 802.1CB, a device can send duplicate frames via separate paths over a multi-hop network \cite{cavalcanti2020802}. 

\begin{figure}[!htbp]
	\centering
	\includegraphics[width=3.5in]{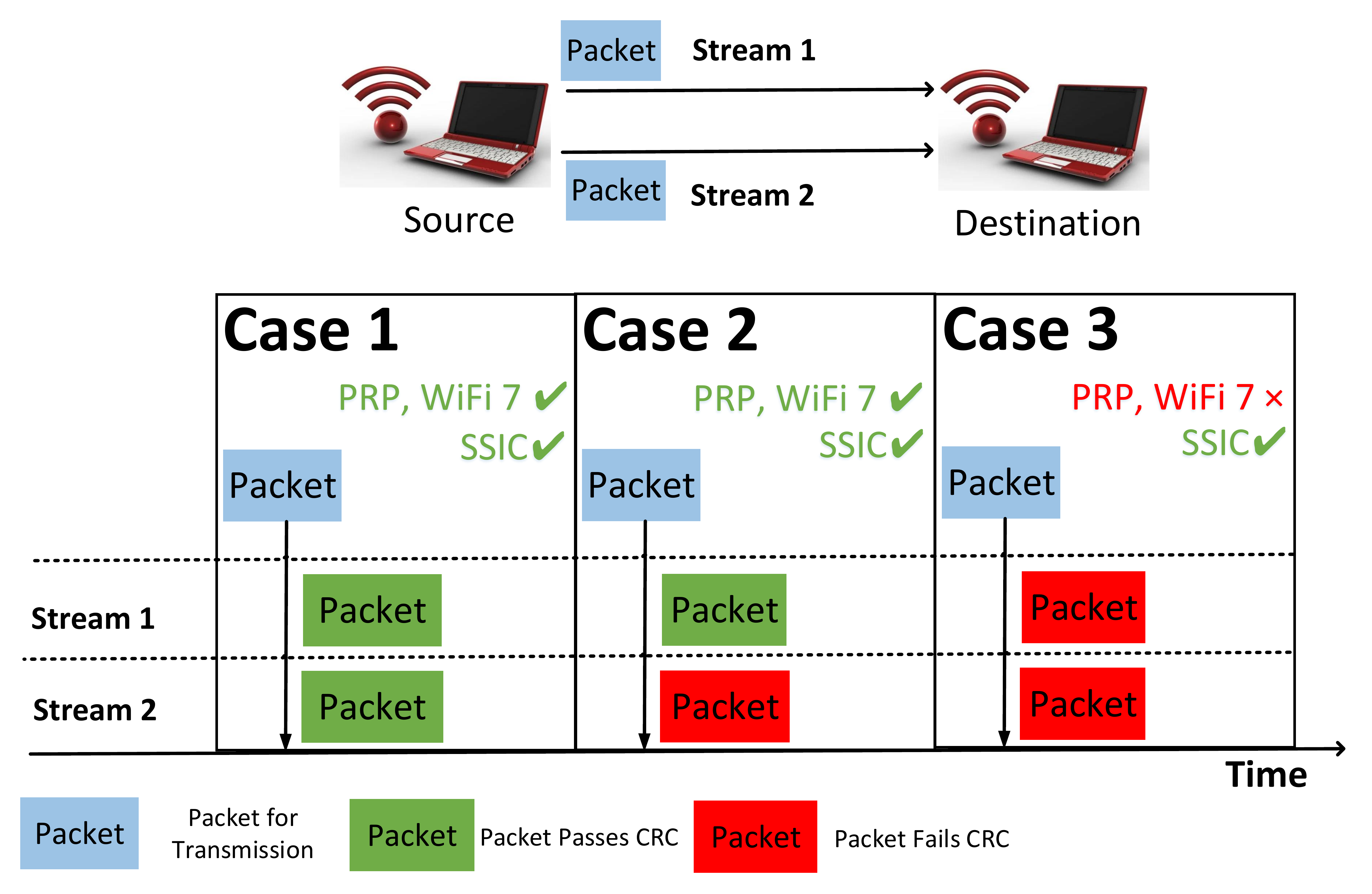}
	\caption{A duplicate transmission setup whereby a source sends two copies of the same packet to a destination over two paths. Three cases are possible: 1) both packets are successfully decoded; 2) only one of the two packets is decoded; 3) both packets fail to be decoded.}
	\label{fig_1}
\end{figure}

In the aforementioned duplicate schemes, duplicates do not undergo joint processing, foregoing a possible way to boost reliability further. Consider a setup in which a source sends two copies of a packet to a destination via two paths, as shown in Fig. \ref{fig_1}. There are three possibilities as far as the reception at the destination is concerned:
\begin{itemize}[]
	\item Case 1: Both packets are decoded successfully, passing the cyclic redundancy check (CRC). 
	\item Case 2: Only one of the packets is decoded, passing the CRC. 
	\item Case 3: Both packets fail to be decoded and do not pass the CRC. 
\end{itemize}

PRP-enabled Wi-Fi and the Wi-Fi 7 proposal work well in Case 2, improving reliability over non-duplicate schemes. However, they do not help in Case 3. Note that failing CRC does not mean having no information on the packet. Consider a packet of 1500 Bytes (the maximum transmission unit of Ethernet). Mis-decoding just one of the 8192 bits still leads to a CRC failure. As observed in \cite{lin2008ziptx}, the percentage of partially correct packets can be up to 55$\%$ in 802.11 networks. Modern channel decoding methods often yield soft information on the source bits that give the log-likelihood ratio (LLR) of the probabilities of the bits being zero and one. 

This paper puts forth a \textbf{\textit{source soft-information combining (SSIC)}} mechanism to exploit the soft information from the decoders. With SSIC, even if both packet copies fail to be decoded (Case 3), combining the soft information of the two copies may lead to a successful decoding. In Case 3 above, if the decoders of the two packets can expose the soft source information for joint signal processing, then there is a chance that the packet can be decoded. However, two challenges need to be addressed before SSIC can be deployed on today’s wireless networks over the TCP/IP networking platform: 

\

\noindent \textbf{Challenge 1: Directly combining the LLRs from the different streams does not work for common wireless systems such as Wi-Fi and WiMAX.} For systems like Wi-Fi and WiMAX, the source, before performing channel coding, scrambles the source bits by masking them (XORing them) with a random binary sequence to avoid long consecutive sequences of $0$s or $1$s. Within a NIC, different random masking sequences are used to XOR successive packets. Also, for the duplicates of the same packet, two NICs at the source may mask the packet with different random sequences. The receiver needs to first decode the mask of each packet copy in order to obtain the unscrambled source bits. For the example in Fig. \ref{fig_1}, the two packets may be scrambled with different masking sequences by the two NICs at the source. The LLRs of the source bits output from the two channel decoders must first be descrambled with their respective scrambling masks before they can be combined. Conventionally, a decoder first converts the LLRs of the whole packets to 0 or 1 binary values, then hard decodes the masking sequence from the preamble part of the binary values. The masking sequence is then used to descramble the $0$ or $1$ values of the subsequent payload (Section \ref{sec:overview:SD} gives more details on the conventional descrambling). This conventional descrambling is not compatible with what we want to do because the channel decoder is essentially performing hard decoding, and soft information is lost. Although one may opt to hard decode the masking sequence only but not the source bits, as will be shown later in this paper, the hard-decoded masking sequence is often erroneous, leading to wrong descrambling of the LLRs, and therefore the wrong combination of the LLRs of the two packets. 

\noindent \textbf{Our Solution:} We devise a new descrambling method, referred to as \textbf{\textit{soft descrambling (SD)}}. First, SD allows the scrambled LLRs of the source bits to be descrambled without being converted to binary values. Hence, SD maintains the soft information required by SSIC. Second, SD also yields the “soft” masking sequence. Descrambling the LLRs using a soft masking sequence (as opposed to a hard masking sequence) may lead to better bit-error-rate (BER) and packet error rate (PER) performance in SSIC. 

\

\noindent \textbf{Challenge 2: Design of an SSIC dispatcher at the source that dispatches the packets over different streams and an SSIC aggregator that combines the packets from different streams into one stream in a way that is compatible with commercial NICs and TCP/IP networking.} Ideally, SSIC should be deployable over today’s TCP/IP networks without the need for specialized NICs, and all unmodified TCP/IP applications should run over the SSIC-enabled network in a transparent manner. Also, other than the need to have multiple physical network paths, it would be desirable not to introduce additional hardware equipment into the system. That is, the overall SSIC networking solution should be compatible with the installed base of legacy networking equipment. 

\noindent \textbf{Our Solution:} We design a TCP/IP-compatible networking called the \textbf{\textit{SSIC networking}} architecture in which the SSIC dispatcher, SSIC aggregator, and SD are all implemented as middleware software running between the MAC layer and IP layer (i.e., a layer 2.5 solution). The middleware creates a virtual connection (VC) between the two end nodes over which the multiply physical paths are deployed. The middleware at the source encapsulates an IP packet into a VC frame, and duplicates of the VC frame are sent over the multiple physical paths. At the receiver, the middleware that serves as the SSIC aggregator performs SD, SSIC, and deduplication (in case multiple copies of the same packet are decoded) before forwarding a single packet copy to the IP layer. Conceptually, with this solution, the IP layers at the two ends run over a “virtual network” created by the middleware. That is, SSIC networking allows multiple streams to be grouped and exposed as a single virtual link to the application for TCP/IP communication without the need to modify legacy TCP/IP applications. 

We note that heterogeneous networking is also possible with the SSIC networking architecture. For example, the two physical paths can be over different types of physical networks (e.g., Wi-Fi and WiMAX). 

To validate the performance of SSIC in a real environment, we set up an SSIC network over Wi-Fi. Experimental results show that SSIC significantly improves the packet delivery success rate in a noisy and lossy wireless environment. Specifically, with two streams over two paths, the SSIC network: i) decreases the PER and packet loss rate (PLR) by more than fourfold compared with a single-stream network; ii) provides 99.99\% reliable packet delivery for short-range communication. Potentially, SSIC could be incorporated into Wi-Fi 7 to improve the performance of its multi-link mode. 

The rest of the paper is organized as follows. Section \ref{sec:overview} gives a quick overview of SSIC, SD for SSIS, and SSIC networking. Section \ref{sec:SystemDesign} delves into the details and nuances of various design issues. Section \ref{sec:SimulationExperiment} presents our simulation results as well as the experimental results over our SSIC network testbed. Section \ref{sec:RelatedWork} discusses the related works and Section \ref{sec:Conclusion} concludes this work.

\section{Overview} \label{sec:overview}
\noindent This section gives an overview of the three key technologies for our system: SSIC, SD for SSIS, and SSIC networking. 

\subsection{SSIC} \label{sec:overview:SSIC}
\begin{figure}[!htbp]
	\includegraphics[width=3.5in]{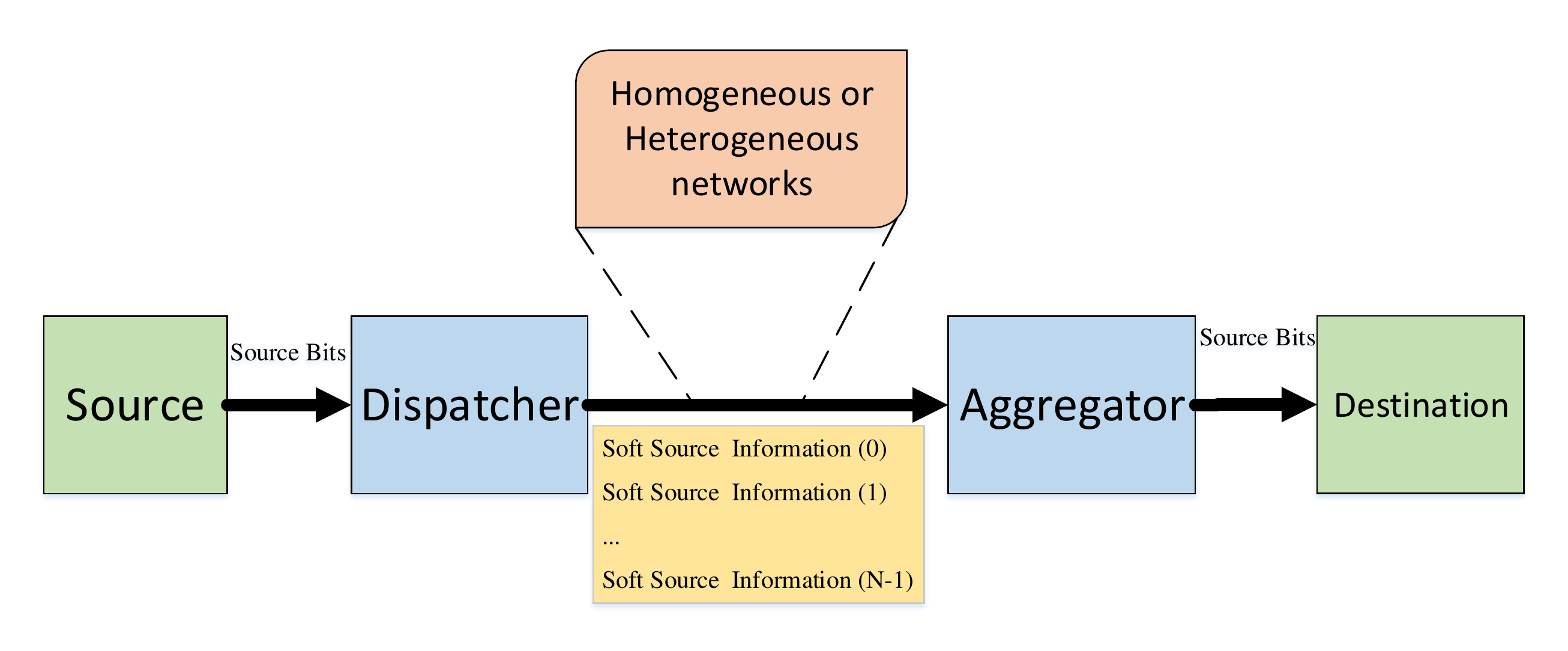}
	\caption{The architecture of SSIC wherein multiple copies of the source information bits are dispatched either over multiple networks of the same kind or multiple networks of different kinds, and wherein the soft bit information from different streams are aggregated for soft combining.}
	\label{fig_2}
\end{figure}

\noindent Fig. \ref{fig_2} shows the architecture of SSIC. The source generates $M$ information bits ${x_0},{x_1},...,{x_{M - 1}}$ and passes them to a dispatcher. The dispatcher sends $N$ copies of information bits over multiple homogeneous networks or multiple heterogeneous networks. The $N$ copies may be subject to different physical-layer signal processing, including the possible use of different modulations and different channel codes. However, even when the decoder of a particular copy $n, n \in \{ 0,1,...,N - 1\}$, cannot decode the source bits ${x_0},{x_2},...,{x_{M - 1}}$, the decoder often has the soft information of the source bits $\tilde x_0^{(n)},\tilde x_1^{(n)},...,\tilde x_{M - 1}^{(n)}$, where $\tilde x_m^{(n)} = \log \left( {{{P[x_m^{(n)} = 0]} \mathord{\left/{\vphantom {{P[x_m^{(n)} = 0]} {P[x_m^{(n)} = 1]}}} \right.\kern-\nulldelimiterspace} {P[x_m^{(n)} = 1]}}} \right)$.

The decoders expose their soft source information to an aggregator. To the extent that the noise corrupting the source information ${x_0},{x_1},...,{x_{M - 1}}$ of different streams are independent, the aggregator can perform soft source information combining as follows:
\begin{equation}
	\label{eqn1}
	\tilde x_m^N \buildrel \Delta \over = \log \left( {{{P[x_m^{} = 0]} \mathord{\left/{\vphantom {{P[x_m^{} = 0]} {P[x_m^{} = 1]}}} \right.\kern-\nulldelimiterspace} {P[x_m^{} = 1]}}} \right) = \sum\limits_{n = 0}^{N - 1} {\tilde x_m^{(n)}}.
\end{equation}
The aggregator decides ${x_m}=0$ if $\tilde x_m^N \ge 0$; and ${x_m}=1$ otherwise. The aggregator may successfully decode ${x_0},{x_1},...,{x_{M - 1}}$ even when none of the $N$ decoders succeed in decoding ${x_0},{x_1},...,{x_{M - 1}}$ from their individual $\tilde x_0^{(n)},\tilde x_1^{(n)},...,\tilde x_{M - 1}^{(n)}.$ 

In general, if decoder $n$ succeeds in decoding ${x_0},{x_2},...,{x_{M - 1}}$, as indicated by its CRC check, it informs the aggregator that there is no need to perform SSIC, and it simply just exposes ${x_0},{x_2},...,{x_{M - 1}}$ to the aggregator. Otherwise, it exposes $\tilde x_0^{(n)},\tilde x_1^{(n)},...,\tilde x_{M - 1}^{(n)}$ to the aggregator. The aggregator performs SSIC only if none of the $N$ decoders succeed in decoding ${x_0},{x_2},...,{x_{M - 1}}$.

\subsection{SD for SSIC} \label{sec:overview:SD}
\begin{figure}[!htbp]
	\centering
	\includegraphics[width=3.6in]{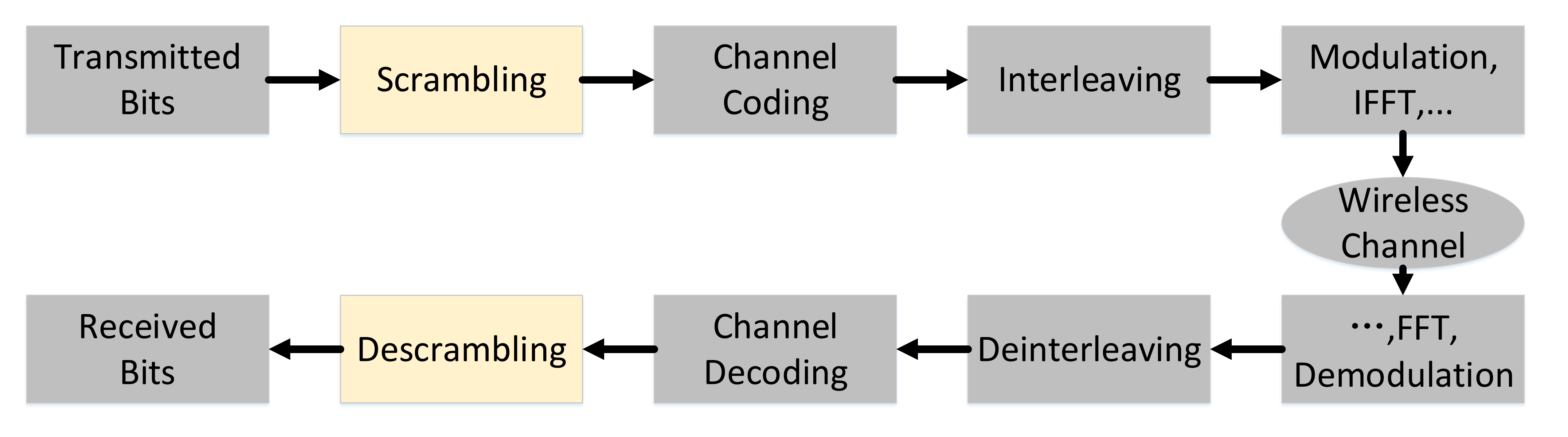}
	\caption{Wi-Fi transceiver’s PHY-layer processing chain.}
	\label{fig_3}
\end{figure}

\noindent In many situations, the source information ${x_0},{x_1},...,{x_{M - 1}}$ is scrambled to avoid long consecutive sequences of $0$s and or $1$s (see \url{https://en.wikipedia.org/wiki/scrambler}). For example, in the Wi-Fi/WiMAX processing chain, as shown in Fig. \ref{fig_3}, the channel-decoded bits at the receiver need to be descrambled in a way to allow the SSIC in \eqref{eqn1} to be performed correctly. 

More specifically, for a stream $n$, what is being transmitted is 
\begin{equation}
	\label{eqn2}
	y_m^{(n)} = {x_m} \oplus z_m^{(n)},{\kern 1pt} {\kern 1pt} {\kern 1pt} m = 0,1,...,M - 1,
\end{equation}
where the scrambling sequence $z_m^{(n)} \in \{ 0,1\}$, $m = 0,1,...,M - 1$, is unique to each stream. As a result, what is obtained from the decoder $n$ is $\tilde y_0^{(n)},\tilde y_1^{(n)},...,\tilde y_{M - 1}^{(n)}$, where $\tilde y_m^{(n)} = \log \left( {{{P[y_m^{(n)} = 0]} \mathord{\left/{\vphantom {{P[y_m^{(n)} = 0]} {P[y_m^{(n)} = 1]}}} \right.\kern-\nulldelimiterspace} {P[y_m^{(n)} = 1]}}} \right)$. To perform SSIC as in \eqref{eqn1}, we need to first descramble $\tilde y_0^{(n)},\tilde y_1^{(n)},...,\tilde y_{M - 1}^{(n)}$ to $\tilde x_0^{(n)},\tilde x_1^{(n)},...,\tilde x_{M - 1}^{(n)}$.

\begin{figure}[!htbp]
	\centering
	\includegraphics[width=2.5in]{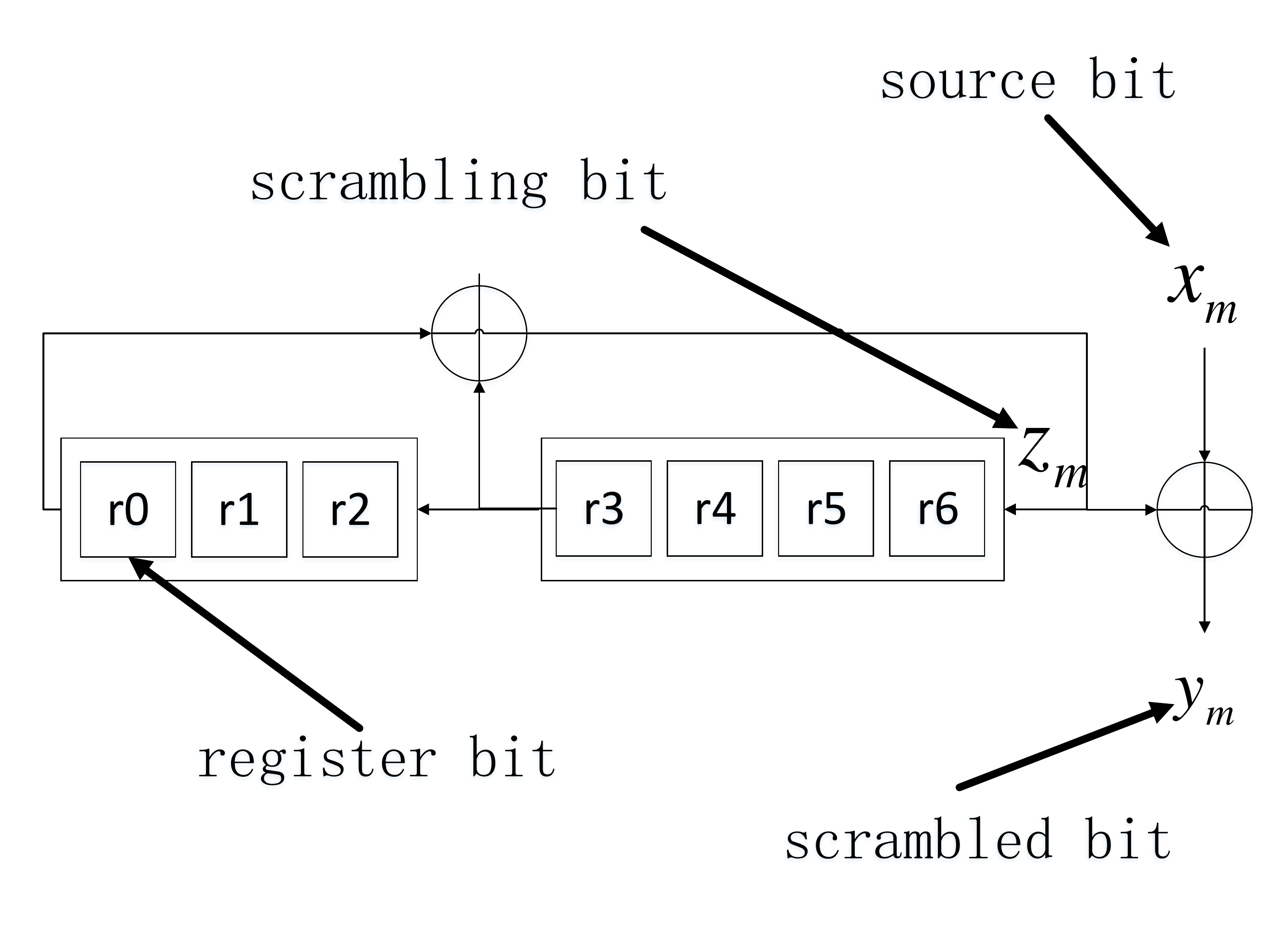}
	\caption{The scrambler in the Wi-Fi system.}
	\label{fig_4}
\end{figure}

Without loss of generality, we explain our idea using a specific scrambler, the $7$-bit and $2$-tap linear-feedback shift register (LFSR) scrambler for Wi-Fi \cite{7786995} shown in Fig. \ref{fig_4}. In the rest of the paper, unless otherwise stated, we focus on an arbitrary decoder out of the $N$ decoders and its associated scrambler and descrambler, and drop the superscript ($n$) in our notations.

Let ${r_{m,j}}$ denote the $j$-th scrambler’s register bit for the generation of ${z_m}$, where $0 \le j \le 6$. The scrambling bit ${z_m}$ is generated by XORing the first and the fourth registers' bits:
\begin{equation}
	\label{eqn3}
	{z_m} = {r_{m,0}} \oplus {r_{m,3}}.
\end{equation}
The scrambled bit ${y_m}$ is derived by
\begin{equation}
	\label{eqn4}
	{y_m} = {z_m} \oplus {x_m} = {r_{m,0}} \oplus {r_{m,3}} \oplus {x_m}.
\end{equation}
Many systems, including Wi-Fi, set the initial seed bits $\{ {r_0},{r_1},...,{r_6}\}$ randomly for each new transmission of a packet. 

At the transmitter side, the seed bits are encoded into the transmitted sequence of bits as follows. Specifically, $L$ pilot bits $\{ {x_{ - L}},{x_{ - (L - 1)}},...,{x_{ - 1}}\} $ are prepended to the data bits $\{ {x_0},{x_1},...,{x_{M - 1}}\} $. The $L$ pilot bits are all set to $0$. Thus, according to the scrambling structure in Fig. \ref{fig_4}, 
\begin{equation}
	\label{eqn5}
	\begin{array}{l}
		y_{-L} = z_{-L} = r_{-L,0} \oplus r_{-L,3} = r_0 \oplus r_3,\\
		{y_{-(L - 1)}} = {z_{-(L - 1)}} = {r_{ - (L - 1),0}} \oplus {r_{ - (L - 1),3}} = {r_1} \oplus {r_4},\\
		...\\
		{y_{ - (L - 6)}} = {z_{ - (L - 6)}} = {r_{ - (L - 6),0}} \oplus {r_{ - (L - 6),3}} = {r_2} \oplus {r_5} \oplus {r_6},\\
		...\\
		{y_{ - 1}} = {h_{ - 1}}({r_0},{r_1},...,{r_6}),
	\end{array}
\end{equation}
where ${h_{ - 1}}({r_0},{r_1},...,{r_6})$ is an XOR of a subset of the seed bits $\{ {r_0},{r_1},...,{r_6}\}$ that depends on the size of $L$. The scrambled sequence, ${y_m}$ for $m = 0,...,M - 1$, depends on the data sequence according to \eqref{eqn4}. Furthermore, in general, ${z_m} = {h_m}({r_0},{r_1},...,{r_6})$ is the XOR of a subset of seed bits $\{ {r_0},{r_1},...,{r_6}\}$.

For illustration, let us first review the conventional descrambling method of Wi-Fi. In the standard Wi-Fi implementation $L=7$ (i.e., there are $7$ seed bits). All the received bits are first hard-decoded into $0$s and $1$s before the descrambling operation. That is, the Wi-Fi channel decoder outputs the estimated value of ${y_m}$: ${\hat y_m} = 0{\rm{ }}$ or ${\rm{ }}1,m =  - L, - (L - 1)...,M - 1$. Suppose that there is no error in the preamble so that ${\hat y_m} = {y_m}, - 7 \le m \le  - 1$. Then,  	
\begin{equation}
	\label{eqn6}
	\begin{array}{l}
		{{\hat y}_{ - 7}} = {y_{ - 7}} = {z_{ - 7}} = {r_0} \oplus {r_3} = {r_{0,0}},\\
		{{\hat y}_{ - 6}} = {y_{ - 6}} = {z_{ - 6}} = {r_1} \oplus {r_4} = {r_{0,1}},\\
		...\\
		{{\hat y}_{ - 1}} = {y_{ - 1}} = {z_{ - 1}} = {r_2} \oplus {r_5} \oplus {r_6} = {r_{0,6}}.
	\end{array}
\end{equation}
Thus, as can be seen from the above, after their estimations, $\{ {\hat y_{ - 7}},{\hat y_{ - 6}},...,{\hat y_{ - 1}}\}  = \{ {z_{ - 7}},{z_{ - 6}},...,{z_{ - 1}}\}$ can be preloaded into the seven registers in the structure in Fig. \ref{fig_4}, to produce the subsequent ${z_m},m = 0,1...,M - 1$. Wi-Fi then descrambles the scrambled source bits ${\hat x_m}, m = 0,1...,M - 1$, by  
\begin{equation}
	\label{eqn7}
	 {\hat x_m} = {\hat y_m} \oplus {z_m},0 \le m \le M - 1.
\end{equation}
Note that the descrambler’s structure is the same as the scrambler’s structure of Fig. \ref{fig_4} except that now ${x_m}$ becomes ${\hat y_m}$, ${y_m}$ becomes ${\hat x_m}$ (i.e., for the descrambler, ${\hat y_m}$ is the input and ${\hat x_m}$ is the output), with ${z_m}$ remaining the same. We refer to this conventional descrambling as \textbf{\textit{hard descrambling (HD)}} since this method uses the hard-bit ${\hat y_m}, - 7\le m\le-1$ to derive the hard-bit ${z_m}$ and the hard-bit ${\hat x_m},0 \le m \le M - 1$. However, HD has two problems:
\begin{enumerate}[label=(\roman*)]
	\item The channel decoder outputs $\{ \hat y_0,\hat y_1,...,\hat y_{M - 1}\} $ rather than $\{\tilde y_0,\tilde y_1,...,\tilde y_{M - 1}\}$. Thus, HD does not yield the soft information $\tilde x_m,0 \le m \le M - 1$ and hence cannot be used  in our proposed SSIC architecture. 
	\item If the decoder fails to decode ${y_m}$ correctly from some $m{\rm{, }} - 7 \le m \le  - 1$, then many of the ${z_m},0 \le m \le M - 1$, and therefore many of the ${x_m},0 \le m \le M - 1$, will be incorrect. 
\end{enumerate}

We propose a \textbf{\textit{soft descrambling (SD)}} technique to circumvent the above problems. Our SD can: i) derive the soft information $\tilde x_m$ needed for SSIC; ii) improve the decoding of $x_m$. Suppose that ${y_m}, - 7 \le m \le M - 1$ is corrupted by noise. Two variants of SD described in the following are possible (see Section \ref{sec:SystemDesign:SD} on the details).

The first variant uses the maximum \textit{a posteriori} (MAP) estimation to estimate the scrambler’s seed bits $\{ {r_0},{r_1},...,{r_6}\}$ by
\begin{equation}
	\label{eqn8}
	\arg {\max _{{r_0},{r_1}...,{r_6}}}P({r_0},{r_1},...,{r_6}|{\tilde y_{ - L}},{\tilde y_{ - (L - 1)}},...,{\tilde y_{ - 1}},S).
\end{equation}
where ${\tilde y_m} = \log \left( {{{P[{y_m} = 0]} \mathord{\left/{\vphantom {{P[{y_m} = 0]} {P[{y_m} = 1]}}} \right.\kern-\nulldelimiterspace} {P[{y_m} = 1]}}} \right), - L \le m - 1$, is the soft information produced by the channel decoder, and $S$ stands for the knowledge of the scrambler structure. Also, note that the MAP estimation is the same as the maximum-likelihood (ML) estimation, $\arg {\max _{{r_0},{r_1}...,{r_6}}}P({\tilde y_{ - L}},{\tilde y_{ - (L - 1)}},...,{\tilde y_{ - 1}}|{r_0},{r_1},...,{r_6},S)$ to the extent that the $127$ non-zero bit patterns of $\{ {r_0},{r_1},...,{r_6}\}$ are equally likely.

Upon deriving $\{ {r_0},{r_1},...,{r_6}\}$, the hard-bit ${z_m}$ can be determined. Then, according to the value of ${z_m}$, ${\tilde x_m}$ is obtained by simply changing the sign of $\tilde y_m$: 
\begin{equation}
	\label{eqn9}
	\tilde x_m^{} = \left\{ {\begin{array}{*{20}{c}}
			{\tilde y_m^{}}&{\text{if } {z_m} = 0}\\
			{ - \tilde y_m^{}}&{\text{if } {z_m} = 1}
	\end{array}} \right..
\end{equation}

The second variant first calculates the probability of ${x_m}$, denoted by $P({x_m})$, as 
\begin{equation}
	\label{eqn10}
	\begin{array}{l}
		P({x_m} = 0) = P({y_m} = 0)P({z_m} = 0) + P({y_m} = 1)P({z_m} = 1),\\
		P({x_m} = 1) = P({y_m} = 1)P({z_m} = 0) + P({y_m} = 0)P({z_m} = 1),
	\end{array}
\end{equation}
where $P({y_m} = 0) = {\log ^{ - 1}}\tilde y_m^{}/(1 + {\log ^{ - 1}}\tilde y_m^{})$ and $P({y_m} = 1) = 1/(1 + {\log ^{ - 1}}\tilde y_m^{})$. The derivation of $P({z_m})$ is as follows. 

First we note that ${z_m}$ is the XOR of a subset of $\{ {r_0},{r_1},...,{r_6}\}$. Let us denote the indexes of the registers in this subset by ${S_m}$. Then $P({z_m})$ is given by 
\begin{equation}
	\label{eqn11}
	\begin{array}{l}
		P({z_m} = 0) \propto \sum\limits_{\scriptstyle\{ {r_0},{r_1},...,{r_6}\} :\hfill\atop
			\scriptstyle\sum\limits_{j \in {S_{m{\kern 1pt} }}} {{r_j}{\kern 1pt} {\kern 1pt} (\bmod 2) = 0} \hfill} {P({r_0},{r_1},...,{r_6}|{{\tilde y}_{ - L}},{{\tilde y}_{ - (L - 1)}},...,{{\tilde y}_{ - 1}},S)} ,\\
		P({z_m} = 1) \propto \sum\limits_{\scriptstyle\{ {r_0},{r_1},...,{r_6}\} :\hfill\atop
			\scriptstyle\sum\limits_{j \in {S_{m{\kern 1pt} }}} {{r_j}{\kern 1pt} {\kern 1pt} (\bmod 2) = 1} \hfill} {P({r_0},{r_1},...,{r_6}|{{\tilde y}_{ - L}},{{\tilde y}_{ - (L - 1)}},...,{{\tilde y}_{ - 1}},S)} .
	\end{array}
\end{equation}
Once $P({z_m})$ is obtained, we use \eqref{eqn10} to obtain $P({x_m} = 0)$ and $P({x_m} = 1)$. And $\tilde x_m^{}$ is calculated by
\begin{equation}
	\label{eqn12}
	\tilde x_m^{} = log{\cfrac{{P({x_m} = 0)}}{{P({x_m} = 1)}}}.
\end{equation}

We have two remarks for these two variants:
\begin{enumerate}[label=(\roman*)]
	\item $L$ in the two variants can be larger than $7$ if desired. As will be shown in Section \ref{sec:SimulationExperiment:simulation}, larger $L$ has better decoding performance because larger $L$ imparts a greater degree of redundancy to the coding of the seed bits $\{ {r_0},{r_1},...,{r_6}\}$ through $y_m, m=-L,...,-1$ to provide more protection for them. 
	\item For the derivation of $\tilde x_m$, the second variant uses the soft information of ${z_m}$ rather than hard-bit ${z_m}$ as in the first variant. As will be shown in Section \ref{sec:SimulationExperiment:simulation}, the second variant, with a better estimate for $\tilde x_m$, outperforms the first variant. 
\end{enumerate}

\subsection{SSIC Networking} \label{sec:overview:NETWORKING}
\noindent SSIC networking overlays a virtual circuit network (VC network) on the Internet. In the VC network, a “virtual connection ID” (VCI) is used to identify a source-destination pair. The SSIC network supports TCP/IP networking among SSIC nodes equipped with multiple NICs. For each SSIC node, all of its MAC addresses, one for each NIC, are associated with a VCI and a single IP address. 

\begin{figure}[!htbp]
	\centering
	
	\subfloat[]{\includegraphics[width=3.5in]{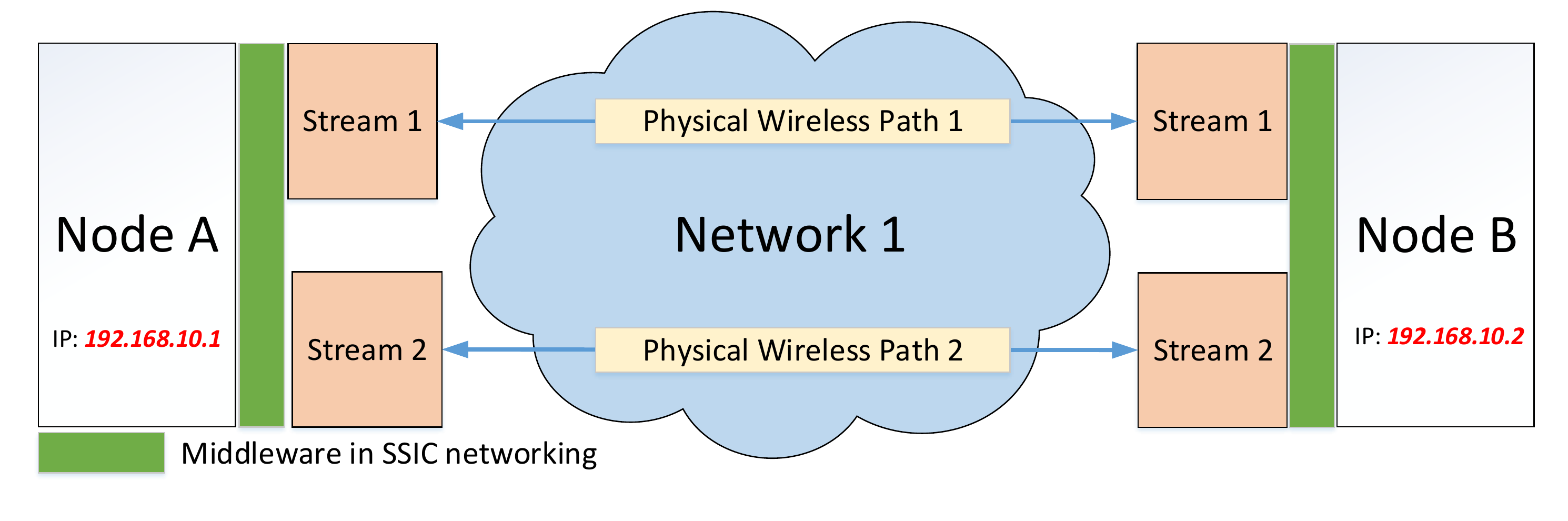}%
	\label{fig_5(a)}}
	
	\subfloat[]{\includegraphics[width=3.5in]{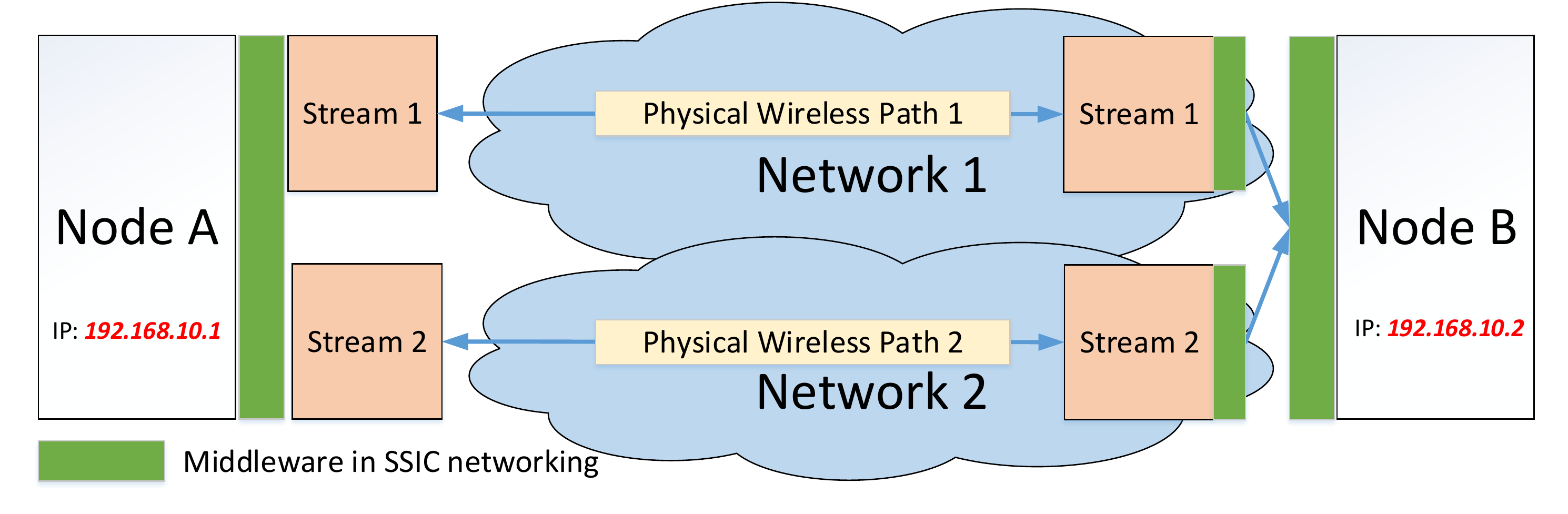}%
	\label{fig_5(b)}}

	\caption{Two multi-homed SSIC nodes, node A and node B, communicate within an SSIC network (192.168.10.0/24) over two separate physical network paths. In (a), the two paths are in the same physical wireless networks, and the two wireless sections end at the same physical locations at both ends. In (b), the two paths are in two different physical wireless networks and the wireless sections of the two paths could end at different physical locations.}
	\label{fig_5}
\end{figure}

Fig. \ref{fig_5} illustrates the main idea. Nodes A and B communicate over using two separate physical wireless paths. These two paths can involve disjoint wireless sections. Particularly, Fig. \ref{fig_5} shows two possible scenarios. In Fig. \ref{fig_5(a)}, the two paths are over the same physical network and terminate at the same end points at node A and at node B. In Fig. \ref{fig_5(b)}, the two paths are over two different physical networks. In this scenario, node A is a mobile device with two NICs to the two networks. Node B is a server within a core network. We will investigate the performance of SSIC in these two scenarios in Section \ref{sec:SimulationExperiment:experiment:1} and Section \ref{sec:SimulationExperiment:experiment:2}.
	
In both two scenarios, a unique VCI, $VC{I_{A,B}}$, is assigned to the VC frames between node A and node B. The IP addresses ${I_A}$ and ${I_B}$ within the same subnet are assigned to node A and node B, respectively. $P_1^A$ and $P_2^A$, node A’s MAC addresses, are associated with $VC{I_{A,B}}$ and ${I_A}$; $P_1^B$ and $P_2^B$, node B’s MAC addresses, are associated with $VC{I_{A,B}}$ and ${I_B}$. This setup will also be used in our experiments in Section \ref{sec:SimulationExperiment:experiment}.

The VC frame is a special Ethernet frame handled by a middleware residing between the network layer and the MAC layer of an SSIC node. We overview the middleware in the following:
\begin{enumerate}[label=(\roman*)]
	\item The middleware works as a dispatcher at the source end. The middleware prepends VCI to a copy of the IP packet to construct a VC frame and dispatches the VC frame to the respective path. 
	\item The middleware works as an aggregator at the destination end. The middleware extracts the IP packet copy (in either hard or soft copy) from each VC frame and aligns different copies of the same IP packet for SSIC (after the application of SD if necessary). The middleware also discards redundant hard copies or soft copies of the same IP packet that arrive after the IP packet has already been successfully decoded. 
\end{enumerate}

Leveraging TUN-device programming \cite{TUNTAP}, we can let the middleware operates at the application layer without the need for hardware implementation. To deploy our SSIC network, the only requirement is to enable the output of a hard or soft packet (that is, the hard bits of a packet when the packet is decoded successfully or the soft information of a packet when the packet is corrupted) to the aggregator. The detailed design of the middleware, as well as the VC frame, will be described in Section \ref{sec:SystemDesign:Networking}.

\section{Detailed System Design} \label{sec:SystemDesign}
\noindent This section delves into the details and nuances of the system design. Section \ref{sec:SystemDesign:SSIC} describes how to acquire the soft information needed for SSIC. Section \ref{sec:SystemDesign:SD} presents two variants of SD variants. Section \ref{sec:SystemDesign:Networking} details the networking issues related to the deployment of SSIC, including the middleware design and the VC frame format. For concreteness, throughout this section, we assume that SSIC is deployed over Wi-Fi networks, although SSIC is a general technique that can be deployed over other networks or heterogeneous networks consisting of networks of different types. 

\subsection{Soft information acquisition} \label{sec:SystemDesign:SSIC}
\noindent Soft information such as ${\tilde x_m}$ or ${\tilde y_m}$ in \eqref{eqn9} can be obtained using a soft-output channel decoder. Take Wi-Fi for example. In the 802.11 standard \cite{7786995}, the LDPC decoder is used to decode LDPC codes, and the hard-output Viterbi decoder is used to decode convolutional codes. For the LDPC coding scheme, the LDPC decoder can readily provide the soft output. For the convolutional coding scheme, we need to replace the conventional hard-output Viterbi decoder with a soft-out decoder, e.g., the soft-output Viterbi algorithm (SOVA) decoder \cite{hagenauer1989viterbi} or the MAP-based BCJR decoder \cite{bahl1974optimal}. In this work, we assume the use of the LDPC coding scheme and the LDPC decoder. 

Given that descrambling is needed in Wi-Fi, the steps for the acquisition of ${\tilde x_m}$ in Wi-Fi are as follows:
\begin{enumerate}[label=(\roman*)]
	\item First, the output of the LDPC decoder $\tilde y_m$ is forwarded to the middleware software.
	\item Then, the SD in the middleware performs descrambling to obtain ${\tilde x_m}$. 
	\item If there are several copies of the ${\tilde x_m}$: $\tilde x_m^n,n = 1,2,3,...,N,$ the middleware then performs SSIC as in \eqref{eqn1}. 
\end{enumerate}

In terms of the local transfer of $\tilde y_m$ from the hardware to the middleware, a simple one-bit field inside the packet informs the middleware whether this packet is hard or soft. Specifically, if it is a hard packet, the packet is a sequence of binary values; otherwise, the packet is a sequence of LLR floating-point values. Meanwhile, a proper quantization method can reduce the storage requirements for LLR values. Much research \cite{volkhausen2012quantization} has been done on the number of bits per floating-point value is needed to reach a certain LLR combining performance. Hence, rather than repeating the investigation here, we adopt a proven quantization method (see Table \ref{tab:2} in Section \ref{sec:SimulationExperiment:experiment}). 

\subsection{SD Methodology} \label{sec:SystemDesign:SD}
\noindent One naïve method to derive ${\tilde x_m}$ from $\tilde y_m^{}$ from the LDPC decoder is as follows:
\begin{enumerate}[label=(\roman*)]
	\item The first seven soft bits $\{ \tilde y_{ - 7}^{},\tilde y_{ - 6}^{},...,\tilde y_{ - 1}^{}\} $ are hard decoded into  $\{ {\hat y_{ - 7}},{\hat y_{ - 6}},...,{\hat y_{ - 1}}\}$. These seven hard bits $\{ {\hat y_{ - 7}},{\hat y_{ - 6}},...,{\hat y_{ - 1}}\}$ are then preloaded into the seven registers in the structure in Fig. \ref{fig_4}.
	\item The scrambler’s seeds $\{ {z_{ - 7}},{z_{ - 6}},...,{z_1}\} $ are obtained using \eqref{eqn6}. With the seeds, all the other ${z_m},m = 0,...,M - 1$ is derived using \eqref{eqn3}.
	\item The soft bits ${\tilde x_m}$ are then obtained by directly changing the sign of $\tilde y_m^{}$ as in \eqref{eqn9} based on the ${z_m}$ obtained in (ii).
\end{enumerate}

Although this method is simple, it is, in fact, the same as the conventional HD as far as the decoding of $\{ {r_0},{r_1},...,{r_6}\}$ is concerned; just that the ${\tilde x_m}$ obtained is still soft information – i.e., this is a hard $\{ {r_0},{r_1},...,{r_6}\}$ soft ${\tilde x_m}$ method. In particular, the system will fail to obtain the correct descrambled ${\tilde x_m}$ if the hard $\{ {r_0},{r_1},...,{r_6}\}$ is mis-decoded. As will be shown in Section \ref{sec:SimulationExperiment:simulation}, this method is far from optimal when deployed with SSIC. We refer to this method as naïve SD since it can still output soft information ${\tilde x_m}$. 

Next we present two advanced variants of SD to obtain better estimated ${\tilde x_m}$. For both variants, redundancy has been introduced to $\{ {r_0},{r_1},...,{r_6}\}$ so that more than seven pilot bits are used to transmit information related to $\{ {r_0},{r_1},...,{r_6}\}$. 

\

\noindent \scalebox{1.14}{\textbf{\textit{First variant: hard-r-soft-x}}:} 

\

The first method is based on the observation that the correct decoding of $\{ {r_0},{r_1},...,{r_6}\}$ is critical. If $\{ {r_0},{r_1},...,{r_6}\}$ is mis-decoded, the soft ${\tilde x_m}$ obtained in the naïve method will not only be useless, but may actually cause harm when ${\tilde x_m}$ is combined with other streams using SSIC because the signs of ${\tilde x_m}$ for many bits will be reversed. For higher protection of $\{ {r_0},{r_1},...,{r_6}\}$, we could introduce redundancy. In particular, instead of using seven pilot bits for $\{ {r_0},{r_1},...,{r_6}\}$, we could use $L \ge 7$ bits to encode $\{ {r_0},{r_1},...,{r_6}\}$. Then, at the receiver, we use the $L$ bits to decode $\{ {r_0},{r_1},...,{r_6}\}$ so as to reduce the error rate. We note that this redundancy is in addition to the PHY-layer redundancy of the convolutional code. The extra redundancy is added in view of the fact that, for SSIC, it is crucial that $\{ {r_0},{r_1},...,{r_6}\}$ from each stream is decoded correctly so that the sign of ${\tilde x_m}$ is not reversed. In this way, aligned combination of ${\tilde x_m}$ from different streams can be performed. 

Recall from Section \ref{sec:overview:SD} that, with $L$ pilot bits, the first variant of SD aims to derive $\{ {r_0},{r_1},...,{r_6}\}$ using the MAP/ML estimation $\arg {\max _{{r_0},{r_1}...,{r_6}}}P({r_0},{r_1},...,{r_6}|{\tilde y_{ - L}},{\tilde y_{ - (L - 1)}},...,{\tilde y_{ - 1}},S)$ as written in \eqref{eqn8}. Let us examine the \textit{a posteriori} probability here. First, let us define $\vec y = ({y_{ - L}},{y_{ - (L - 1)}},...,{y_{ - 1}})$ and $\vec r = ({r_0},{r_1},...,{r_6})$. For $L \ge 7$, there is a one-to-one mapping from $\vec r$ to $\vec y$ (i.e., not all bit patterns of $\vec y$ are possible; only $127$ of them are possible, each corresponding to a particular $\vec r$). We can write $\vec y$ as a function of $\vec r$ as $\vec y(\vec r)$. In fact, for a given $\vec r$, ${y_m}, - L \le m \le  - 1$, are all determined. Thus, we can also write ${y_m}$ as a function of $\vec r$ as ${y_m}(\vec r)$. Then, the \textit{a posteriori} probability $P({r_0},{r_1},...,{r_6}|{\tilde y_{ - L}},{\tilde y_{ - (L - 1)}},...,{\tilde y_{ - 1}},S)$ can be written as 

\begin{equation}
	\label{eqn13}
	\begin{array}{l}
		P(\vec r|{{\tilde y}_{ - L}},{{\tilde y}_{ - (L - 1)}},...,{{\tilde y}_{ - 1}},S)\\
		= P({y_{ - L}}(\vec r){y_{ - (L - 1)}}(\vec r)...{y_{ - 1}}(\vec r)|{{\tilde y}_{ - L}},{{\tilde y}_{ - (L - 1)}},...,{{\tilde y}_{ - 1}},S){\kern 1pt} \\
		= \cfrac{P({y_{ - L}}(\vec r)|{{\tilde y}_{ - L}})...P({y_{ - 1}}(\vec r)|{{\tilde y}_{ - 1}})} { {\sum\limits_{\vec r \in \vec R} {P({y_{ - L}}(\vec r)|{{\tilde y}_{ - L}})...P({y_{ - 1}}(\vec r)|{{\tilde y}_{ - 1}})}}}
	\end{array}
\end{equation}
where $\vec R$ is the set of the $127$ non-zero bit patterns for $\{ {r_0},{r_1},...,{r_6}\} $. Define 
\begin{equation}
	\label{eqn14}
	f(\vec r) = {f_{ - L}}(\vec r)...{f_{ - 1}}(\vec r),
\end{equation}
where
\begin{equation}
	\label{eqn15}
	{f_m}(\vec r) = P({y_m}(\vec r)|{\tilde y_m}), - L \le m \le  - 1.
\end{equation}
We can then write the \textit{a posteriori} probability as 
\begin{equation}
	\label{eqn16}
	P(\vec r|{\tilde y_{ - L}},{\tilde y_{ - (L - 1)}},...,{\tilde y_{ - 1}},S) = \cfrac{{f(\vec r)}}{{\sum\limits_{\vec r \in \vec R} {f(\vec r)} }}
\end{equation}
and
\begin{equation}
	\label{eqn17}
	\mathop {\arg \max }\limits_{\vec r} P(\vec r|{\tilde y_{ - L}},{\tilde y_{ - (L - 1)}},...,{\tilde y_{ - 1}},S) = \mathop {\arg \max }\limits_{\vec r} f(\vec r)
\end{equation}

\

\noindent \textbf{\textit{Deriving} ${y_m}(\vec r), - L \le m \le  - 1$}:

\

We note that ${y_m}(\vec r)$ is the XOR of a subset of the seed bits $\{ {r_0},{r_1},...,{r_6}\}$ that depends on the scrambler structure. The assembly of the   equations can be written in matrix form as 
\begin{equation}
	\label{eqn18}
	{\left[ {\begin{array}{*{20}{c}}
				{{y_{ - L}}}&{{y_{ - (L - 1)}}}&{...}&{{y_{ - 1}}}
		\end{array}} \right]^T} = A{\left[ {\begin{array}{*{20}{c}}
				{{r_0}}&{{r_1}}&{...}&{{r_6}}
		\end{array}} \right]^T}
\end{equation}
where $A = \left[ {\begin{array}{*{20}{c}}
		{{a_{ - L,0}}}&{{a_{ - L,1}}}& \ldots &{{a_{ - L,6}}}\\
		{{a_{ - (L - 1),0}}}&{{a_{ - (L - 1),1}}}& \ldots &{{a_{ - (L - 1),6}}}\\
		\vdots & \vdots & \ddots & \vdots \\
		{{a_{ - 1,0}}}&{{a_{ - 1,1}}}& \ldots &{{a_{ - 1,6}}}
\end{array}} \right]$ and ${a_{m,j}} = 1$ or ${\rm{ }}0, - L \le m \le  - 1,0 \le j \le 6$. Note that the summation operation in \eqref{eqn18} is a mod-two operation. Thus, for each ${y_m}$ we have
\begin{equation}
	\label{eqn19}
	{y_m}(\vec r) = {a_{m,0}}{r_0} \oplus {a_{m + 1,1}}{r_1}... \oplus {a_{m + L - 1,6}}{r_6}.
\end{equation}

We can devise an algorithm to determine the value of ${a_{m,j}}$, which depends on the scrambler structure, as follows (in the following, ${S_m}$ is the indexes of the subset of the registers whose XOR produce ${y_m}$, i.e., ${y_m} = \mathop {\sum {} }\limits_{i \in {S_m}} {r_i} (\bmod 2)$):
\begin{enumerate}[label=(\roman*)]
	\item Initialize seven sets: ${S_{ - L - 7}} = \{ 0\} $, ${S_{ - L - 6}} = \{ {\rm{1}}\} $, ${S_{ - L - 5}} = \{ {\rm{2}}\} $, ${S_{ - L - 4}} = \{ {\rm{3}}\} $, ${S_{ - L - 3}} = \{ {\rm{4}}\} $, ${S_{ - L - 2}} = \{ {\rm{5}}\} $ and ${S_{ - L - 1}} = \{ {\rm{6}}\} $. 
	\item For each $m, - L\le m \le -1$, we further define a corresponding set ${S_m}$ as:
	\begin{equation}
		\label{eqn20}
		{S_m} = {S_{m - 7}} \cup {S_{m - 4}} - {S_{m - 7}} \cap {S_{m - 4}},
	\end{equation}
	where the set operations in \eqref{eqn20} realize the XOR operation that produce the ${y_m}$ (i.e., ${z_m}$ given that the pilot bits ${x_m}$ are $0$) in the scrambler structure of Fig. \ref{fig_4}. 
	\item Given a specific $m, - L \le m \le  - 1$ and $j,0 \le j \le 6$, ${a_{m,j}}$ is determined by
	\begin{equation}
	\label{eqn21}
	{a_{m,j}} = \left\{ {\begin{array}{*{20}{c}}
				1,&{\text{if } j \in {S_m}}\\
				0,&{\text{if } j \notin {S_m}}
	\end{array}} \right..
	\end{equation}
\end{enumerate}
After determining ${a_{m,j}}$, we can use \eqref{eqn19} to list the $L$ equations of ${y_m}$ and $\{ {r_0},{r_1}...,{r_6}\} $. Note that the derivation of ${a_{m,j}}$, which is independent of the values of $\{ {r_0},{r_1}...,{r_6}\} $, only needs to run once during initialization. 

\

\noindent \textbf{\textit{Deriving ${f_m}(\vec r), - L \le m \le  - 1,$ and $f(\vec r)$}}:

\

From \eqref{eqn15}, we have 
\begin{equation}
	\label{eqn22}
	{f_m}(\vec r) = P({y_m}(\vec r)|{\tilde y_m}) = \left\{ {\begin{array}{*{20}{c}}
			{\cfrac{{{{\log }^{ - 1}}\tilde y_m^{}}}{{1 + {{\log }^{ - 1}}\tilde y_m^{}}},}&{\text{if }{y_m}(\vec r) = 0}\\
			{\cfrac{1}{{1 + {{\log }^{ - 1}}\tilde y_m^{}}},}&{\text{if }{y_m}(\vec r) = 1}
	\end{array}} \right.
\end{equation}
where $\tilde y_m^{} = \log \left( {{{P[{y_m} = 0]} \mathord{\left/{\vphantom {{P[{y_m} = 0]} {P[{y_m} = 1]}}} \right.\kern-\nulldelimiterspace} {P[{y_m} = 1]}}} \right)$ is the output of the LDPC decoder as discussed in Section \ref{sec:SystemDesign:SSIC}. The $L$ values of ${f_m}(\vec r), - L \le {\rm{ }}m \le  - 1$ obtained by \eqref{eqn22} can then be substituted into \eqref{eqn14} to get $f(\vec r)$.

With $f(\vec r)$, the first variant is as follows: 
\begin{enumerate}[label=(\roman*)]
	\item From \eqref{eqn17}, we obtain the maximum \textit{a posteriori} ${\vec r^*} = \mathop {\arg \max }\limits_{\vec r} f(\vec r)$.
	\item We load ${\vec r^*}$ as the initial values to the scrambler’s registers, and ${z_m},m \ge 0,$ can then be obtained as the successive outputs form the scrambler. Then, according to the value of  ${z_m}$, ${\tilde x_m}$ is obtained by simply changing the sign of $\tilde y_m$ using \eqref{eqn9}.
\end{enumerate}

We refer to this variant as the \textbf{\textit{hard-r-soft-x (HRSX)}} since it hard decodes the scrambling bit ${z_m}$ (i.e., we hard decode $\{ {r_0},{r_1}...,{r_6}\} $ which gives rise to hard ${z_m}$) to derive the soft information ${\tilde x_m}$. In the meantime, it is worthwhile to note that the naïve method is also a kind of HRSX, but without the “redundant encoding” of ${\tilde x_m}$ in ${y_{ - L}},{y_{ - (L - 1)}},...,{y_{ - 1}}$.

\

\noindent \scalebox{1.14}{\textbf{\textit{Second variant: soft-r-soft-x}}:}

\

\noindent The second variant uses the estimated probability of ${z_m}$ to improve the estimation. Note that for the estimated probability of ${x_m}$, $P({x_m})$, we have 
\begin{equation}
	\label{eqn23}
	\begin{array}{l}
		P({x_m}=0) = P({y_m}=0)P({z_m}=0)+ P({y_m}=1)P({z_m}=1),\\
		P({x_m}=1) = P({y_m}=1)P({z_m}=0)+ P({y_m}=0)P({z_m}=1).
	\end{array}
\end{equation}
We have \eqref{eqn23} because the noises incurred in obtaining the probabilities $P({y_m})$ and $P({z_m})$ are also independent. Once we obtain the above $P({x_m})$ for ${x_m}$, we can also obtain ${\tilde x_m}$. 

The following describes the steps of this variant:
\begin{enumerate}[label=(\roman*)]
	\item First, $f(\vec r)$ for $- L \le m \le  - 1$ is derived as in \textbf{\textit{HRSX}}.
	\item Note that ${z_m}$ is the XOR of a subset of the possible values for $\vec r$ and has a period of $127$. Let ${\vec R_m} = \{ {r_0},{r_1},...,{r_6} : \sum\limits_{j \in {S_m}} {{r_j}} {\kern 1pt} {\kern 1pt} {\kern 1pt} {\kern 1pt} {\kern 1pt} (\bmod 2) = 0)\} $ where the summation here is the mod $2$ summation. The $P({z_m})$ for $- L \le m \le M - 1$ is given by
	\begin{equation}
		\label{eqn24}
		\begin{array}{l}
		P({z_m} = 0) = \cfrac{{\sum\limits_{\vec r \in {{\vec R}_m}} {f(\vec r)} }}{{\sum\limits_{\vec r 	\in \vec R} {f(\vec r)} }},\\
		P({z_m} = 1) = 1 - P({z_m} = 0).
		\end{array}
	\end{equation}
	\item Once $P({z_m})$ is derived, we obtain $P({x_m})$ using \eqref{eqn23}. Then, ${\tilde x_m}$ is given by 
	\begin{equation}
	\label{eqn25}
	{\tilde x_m} = \log \left( {{{P[x_m^{} = 0]} \mathord{\left/
				{\vphantom {{P[x_m^{} = 0]} {P[x_m^{} = 1]}}} \right.
				\kern-\nulldelimiterspace} {P[x_m^{} = 1]}}} \right)
	\end{equation}
\end{enumerate}
We refer to the second variant as the \textbf{\textit{soft-r-soft-x (SRSX)}} since it soft decodes the scrambling bit ${z_m}$ to derive ${\tilde x_m}$.

In conclusion, all variants of SD, including the naïve SD (HRSX without redundancy), HRSX, and SRSX, can be deployed in the SSIC network. Specifically, given $N,N \ge 1$ streams in total, after deriving $\tilde x_m^{}$ by using any variant of SD, the information bit $x_m^{}$ is finally decoded by 
\begin{equation}
	\label{eqn26}
	x_m = \left\{ {\begin{array}{*{20}{c}}
			0, &{\text{if }\tilde x_m^N = \sum\limits_{n = 0}^{N - 1} {\tilde x_m^{(n)}}  \ge 0}\\
			1, &{\text{if }\tilde x_m^N{\rm{ = }}\sum\limits_{n = 0}^{N - 1} {\tilde x_m^{(n)}}  < 0}
	\end{array}} \right..
\end{equation}

\subsection{Detailed design of SSIC networking} \label{sec:SystemDesign:Networking}
\begin{figure}[!htbp]
	\centering
	\includegraphics[width=2.6in]{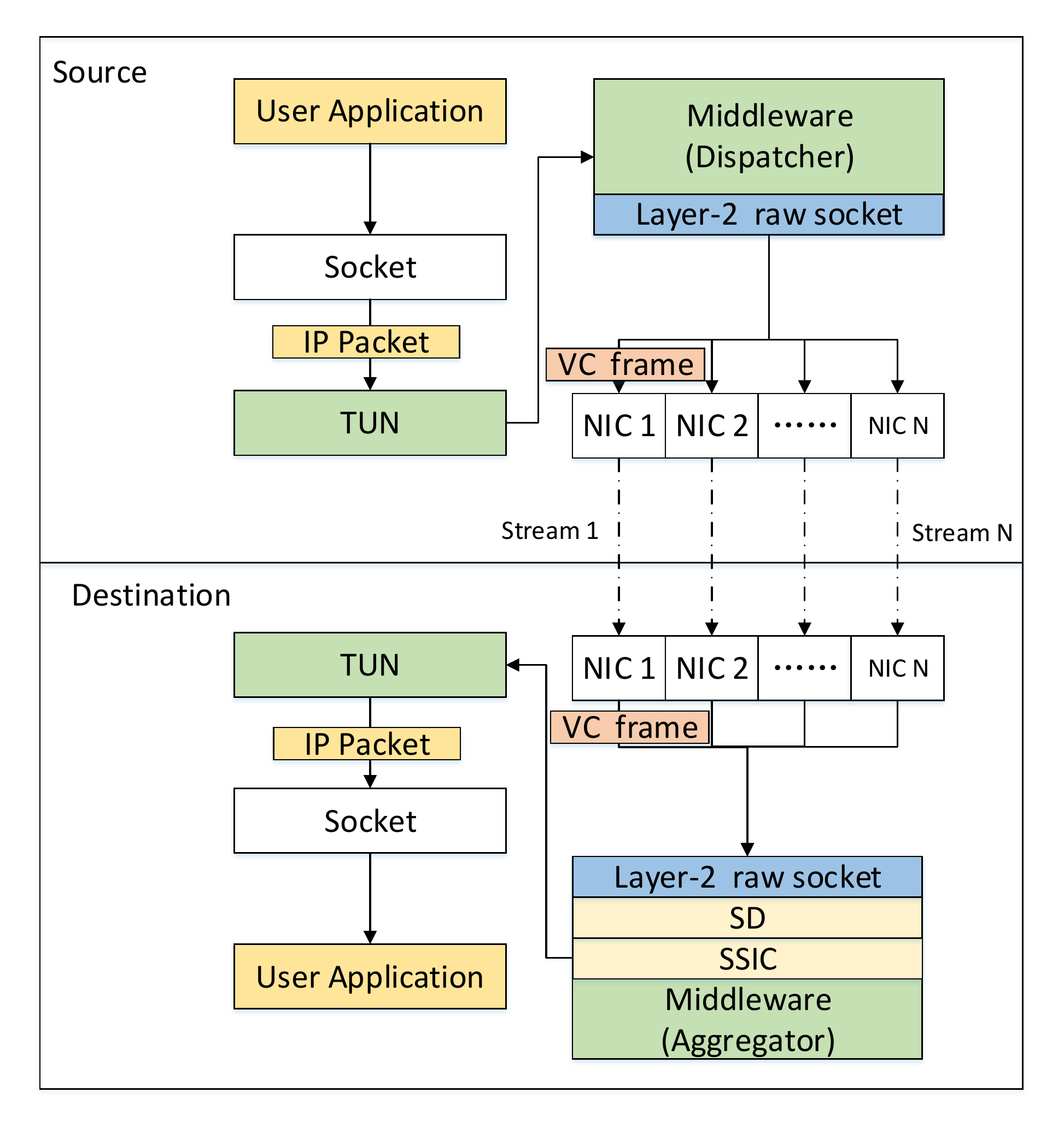}
	\caption{The detailed working mechanism of the middleware.}
	\label{fig_7}
\end{figure}

\noindent This subsection details the SSIC networking framework, including the middleware design and the VC frame format. As illustrated in Fig. \ref{fig_7}, a TUN device \cite{TUNTAP} is first created along with the middleware on each SSID node. The IP address of each node’s TUN device is set to the VC IP address. The user application, on the other hand, can use any type of socket (e.g., TCP socket, UDP socket, or raw socket) to use the SSIC network service. A packet generated from the user application is sent to the network stack by the socket. Thanks to the TUN devices in the operation system (OS), if the destination IP address is located within the SSIC network, this packet is then forwarded to the TUN device by the network stack according to the system’s routing table. 

\begin{figure}[!htbp]
	\centering
	\includegraphics[width=3.5in]{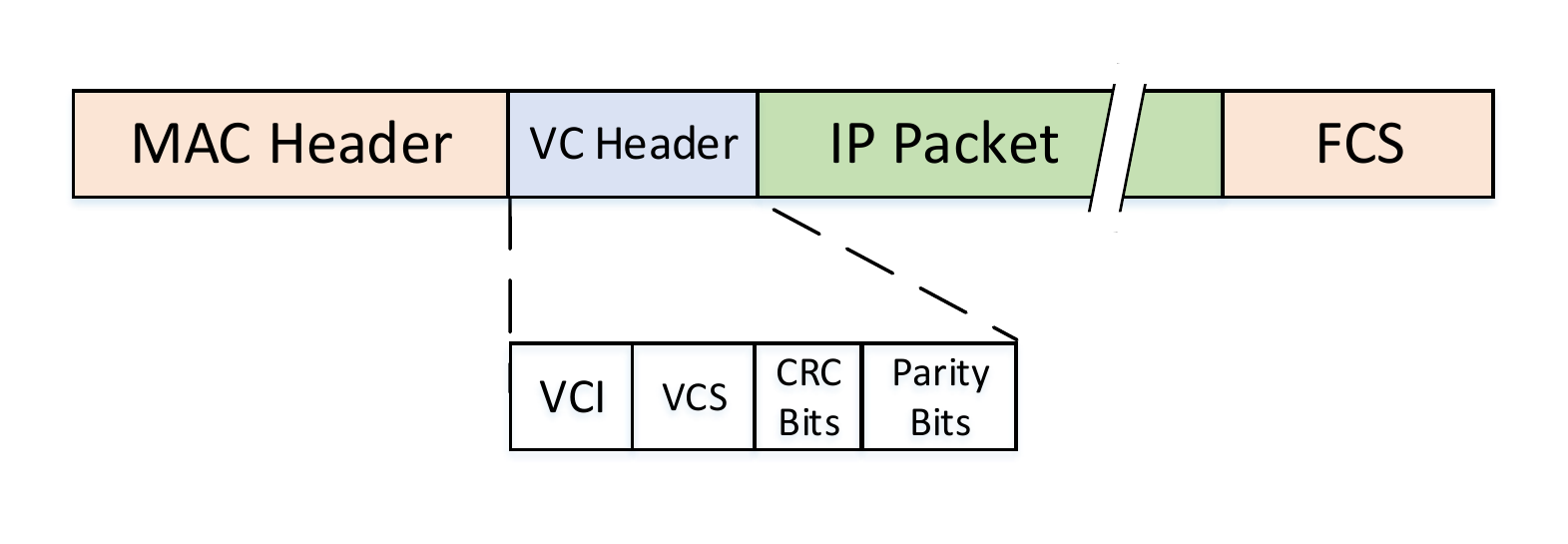}
	\caption{The layer-2 VC frame format: A VC header is introduced in front of the layer-2 payload.}
	\label{fig_8}
\end{figure}

The VC frame format is shown in Fig. \ref{fig_8}, where a VC header is added in between the MAC header and a copy of the IP packet. The VC header includes a VCI to identify the source-destination pair, a VC sequence number (VCS) to identify and match the copies of an IP packet, CRC bits, and parity bits. In particular, the parity bits are to provide extra-strong protection to the header, making sure that the header can be decoded to hard bits correctly (by checking the CRC bits) even if the receiver’s VC frame fails the layer-2 CRC. With the VC being identified by the decoded header, at the destination end, SSIC can be performed over the VC frames with the same VC header to decode the IP packet. 

At the source end, an IP packet with ${I_A}$ as source IP and ${I_B}$ as destination IP is generated by the user application. Then the IP packet is forwarded to TUN and read by the middleware. Upon getting the IP packet from TUN, the middleware encapsulates the IP packet into a VC frame. After generating a VC frame, the middleware sends it to its corresponding NIC using a layer-2 raw socket. 

\

\noindent \textbf{\textit{Remark 1: At the source end, the middleware generates multiple VC frames carrying the same IP packet and the same VC header. However, VC frames destined for different NICs have different MAC headers.}}

\

At the destination end, the middleware receives either a hard or a soft VC frame carrying an IP packet from any one of the NICs. The MAC layer returns the hard VC frame if it passes the layer-2 CRC; or the soft VC frame if it fails the layer-2 CRC. Upon receiving a VC frame, the middleware performs the following steps: 
\begin{enumerate}[label=(\roman*)]
	\item If the VC frame is a hard frame, the middleware directly goes to the final step.
	\item If the VC frame is a soft frame, the middleware first performs SD if the scrambling process exists in the lower layer. Then it hard-decodes the VCI, VCS, and CRC from the VC header and checks its CRC status. If the frame passes CRC, the middleware detects a valid VC frame and goes to the next step; otherwise, the VC frame is discarded.
	\item The middleware checks whether it has received the same VC frame that passes the layer-2 CRC before. If yes, the latest received VC frame is also discarded as a duplicate; otherwise, it goes to the next step.
	\item The middleware checks whether it has received a soft VC frame with the same VCI and VCS as that of the latest VC frame. If yes, the middleware collects all these VC frames and performs SSIC; otherwise, the latest soft VC frame is stored locally in the middleware.
	\item The middleware extracts the IP packet from the VC frame and writes it to the TUN device. From there, the IP packet goes to the user application. The reception of the IP packet is then complete. 
\end{enumerate}

\

\noindent \textbf{\textit{Remark 2: On each SSIC node, the MAC layer returns soft VC frames when the VC frame fails the layer-2 CRC so that SD and SSIC can be performed.}}

\section{Simulation and Experimental Validations} \label{sec:SimulationExperiment}
\noindent Section \ref{sec:SimulationExperiment:simulation} presents simulation results to validate the theoretical performance of SD, Section \ref{sec:SimulationExperiment:experiment} presents experimental results over an SSIC network in a real environment.
 
\subsection{Theoretical Performance of SD} \label{sec:SimulationExperiment:simulation}
\noindent To validate the theoretical performance of SD, we performed simulations in MATLAB in which the WLAN format waveform was generated and passed through the AWGN channel. The detailed settings are listed in Table \ref{tab:1}. In this simulation, four streams with different settings were used. Specifically, stream $n,1 \le n \le 4$, has an SNR of $(0.5(n - 1) + p)$ dB at the index of $p,8 \le p \le 16$. That is, among all these four streams, stream $1$ experiences the lowest SNR throughout the experiment, whereas stream $4$ experiences the largest SNR. For example, stream $1$ has an SNR of $8$ dB at index $8$, and stream $4$ has an SNR of $17.5$ dB at index $16$. 

\begin{table}[!htbp]
	\caption{Simulation settings}
	\label{tab:1}
	\begin{tabular}{|l|l|lll}
		\cline{1-2}
		Payload   Size                         & 1500   Bytes                                 &  &  &  \\ \cline{1-2}
		Modulation   and Channel Coding scheme & 64   QAM, 2/3 Code Rate (LDPC)               &  &  &  \\ \cline{1-2}
		Number   of steams                     & 4                                            &  &  &  \\ \cline{1-2}
		Steam   1 SNR range                    & 8dB   to 18dB                                &  &  &  \\ \cline{1-2}
		Steam   2 SNR range                    & 8.5dB   to 18.5dB                            &  &  &  \\ \cline{1-2}
		Stream 3 SNR range                     & 9dB   to 19dB                                &  &  &  \\ \cline{1-2}
		Stream 4 SNR range                     & 9.5dB   to 19.5dB                            &  &  &  \\ \cline{1-2}
	\end{tabular}
\end{table}

\begin{figure}[!htbp]
	\centering
	\subfloat[]{\includegraphics[width=2.3in]{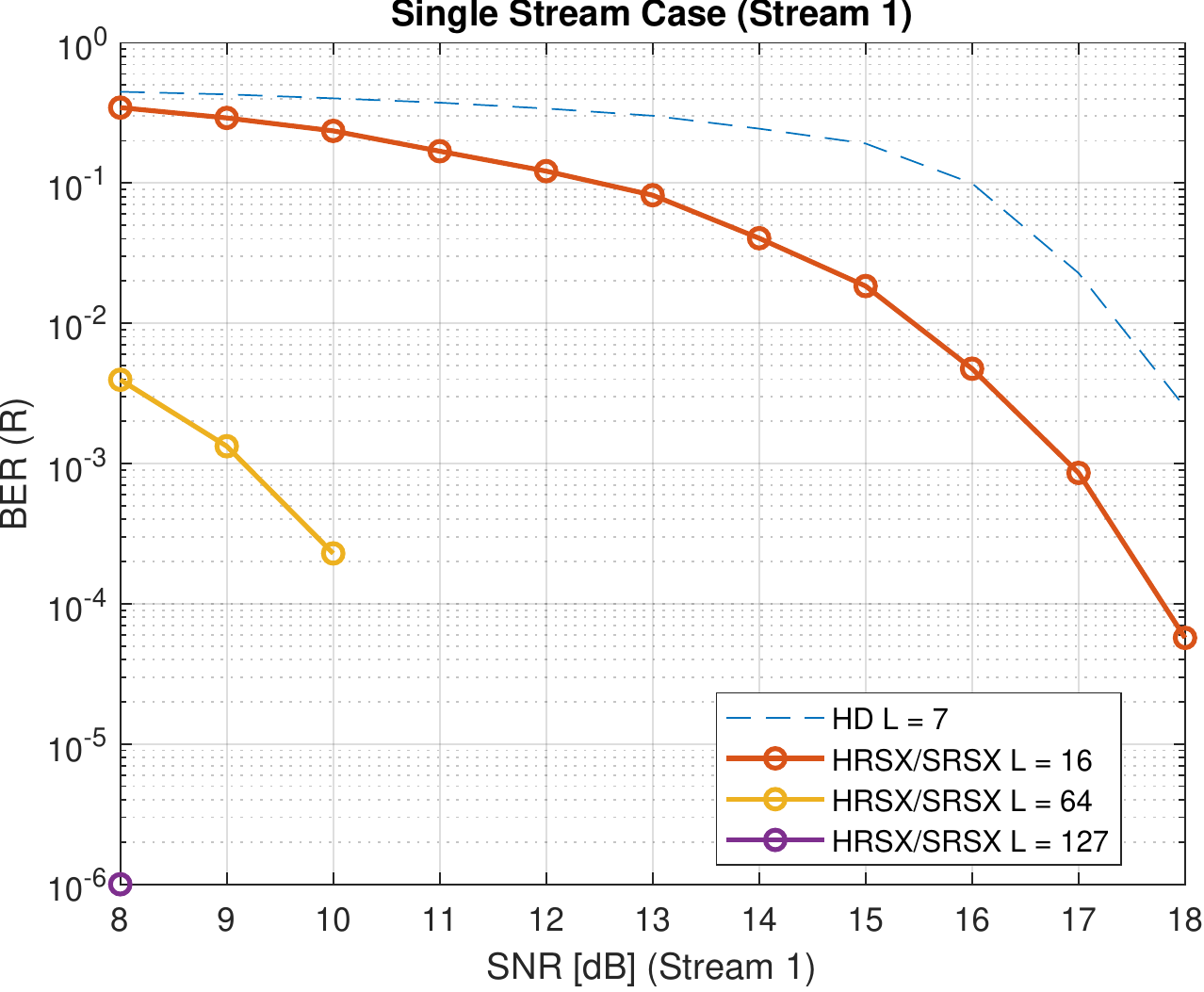}%
		\label{fig_9(a)}}
	
	\subfloat[]{\includegraphics[width=2.3in]{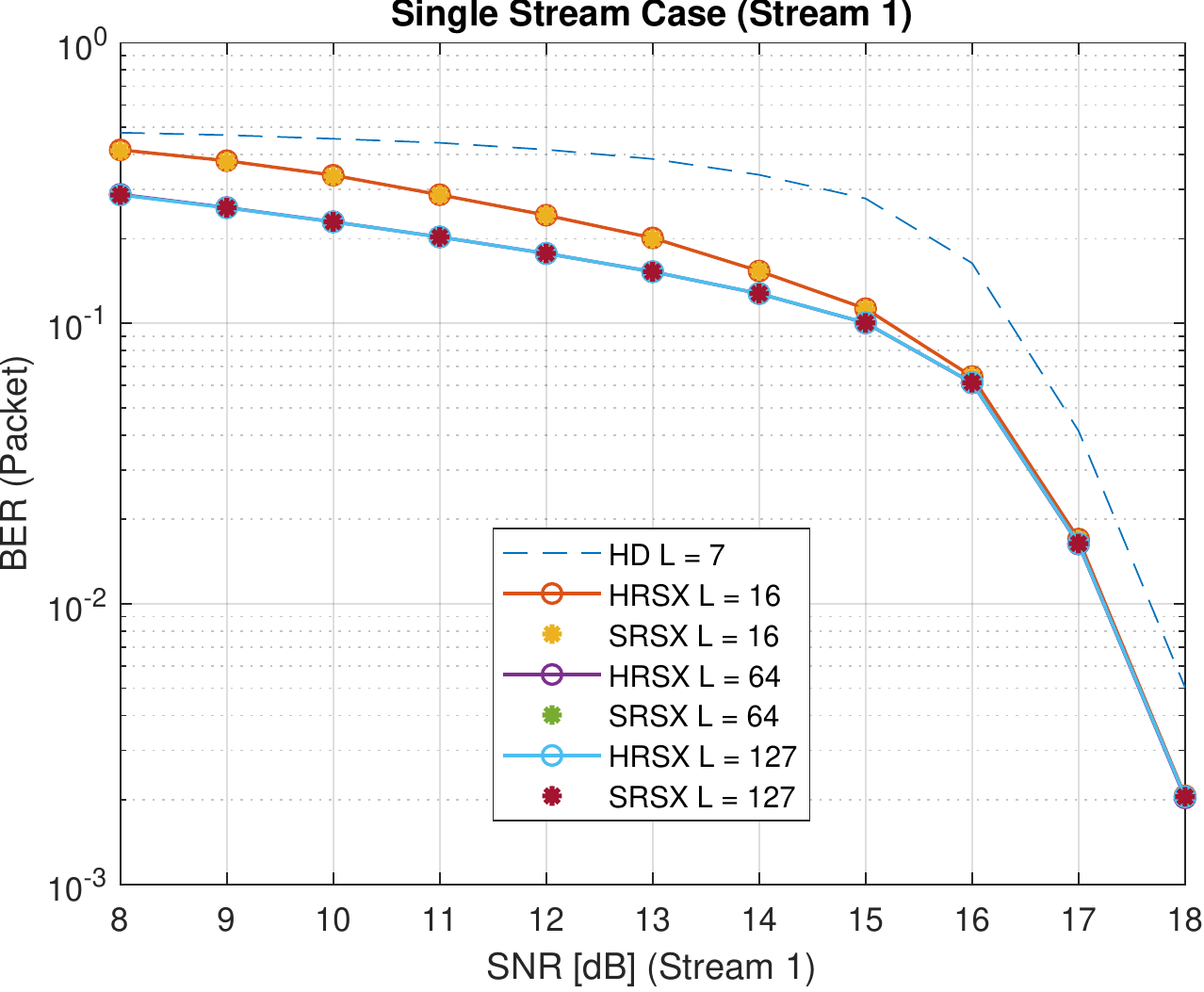}%
		\label{fig_9(b)}}
	\caption{BER of the single-stream case with different $L$: (a) seed-bit decoding performance of HD, HRSX and SRSX; (b) payload decoding performance of HD, HRSX, and SRSX is investigated.}
	\label{fig_9}
\end{figure}

\begin{figure}[!htbp]
	\centering
	\subfloat[]{\includegraphics[width=2.3in]{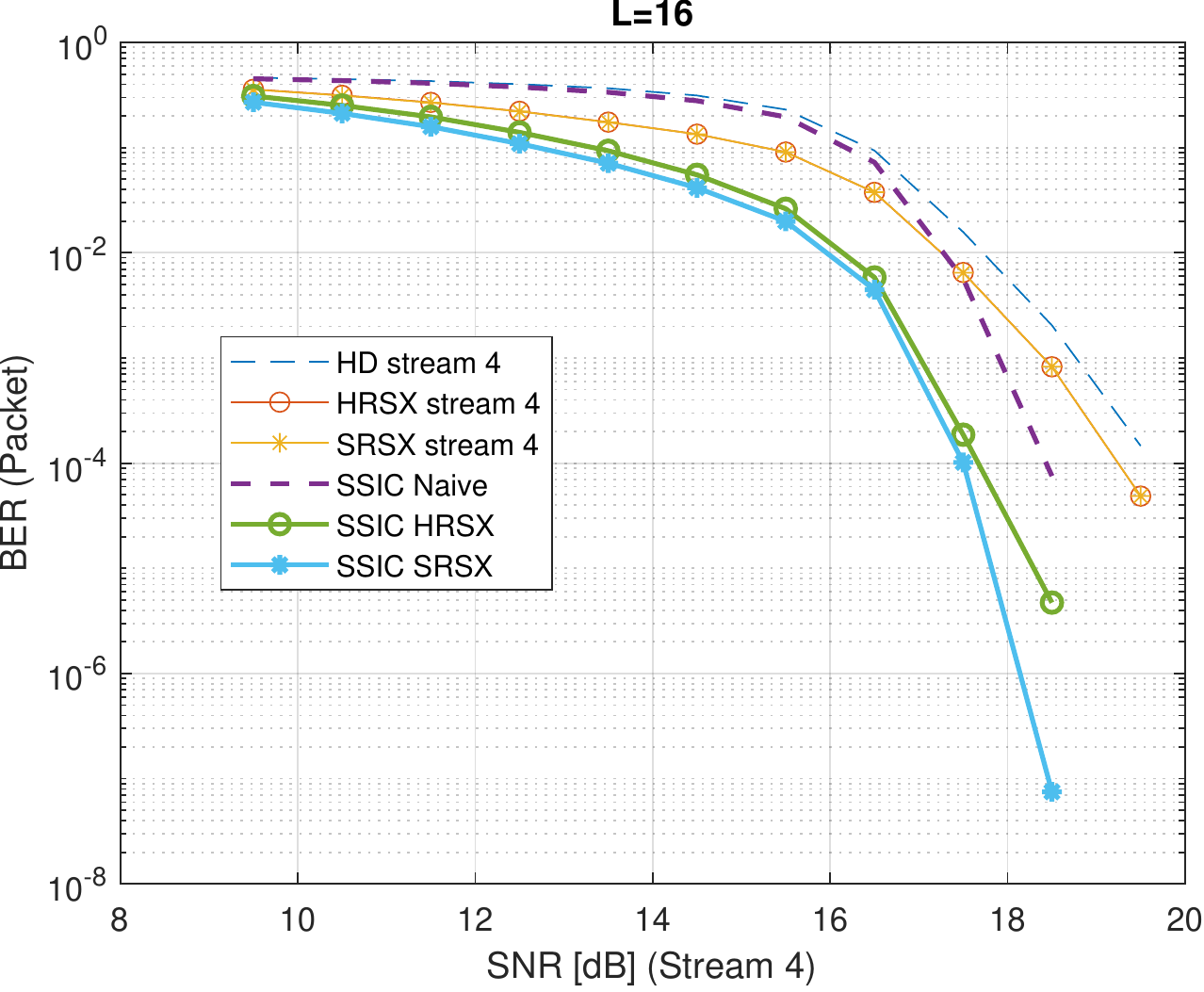}%
		\label{fig_10(a)}}
	
	\subfloat[]{\includegraphics[width=2.3in]{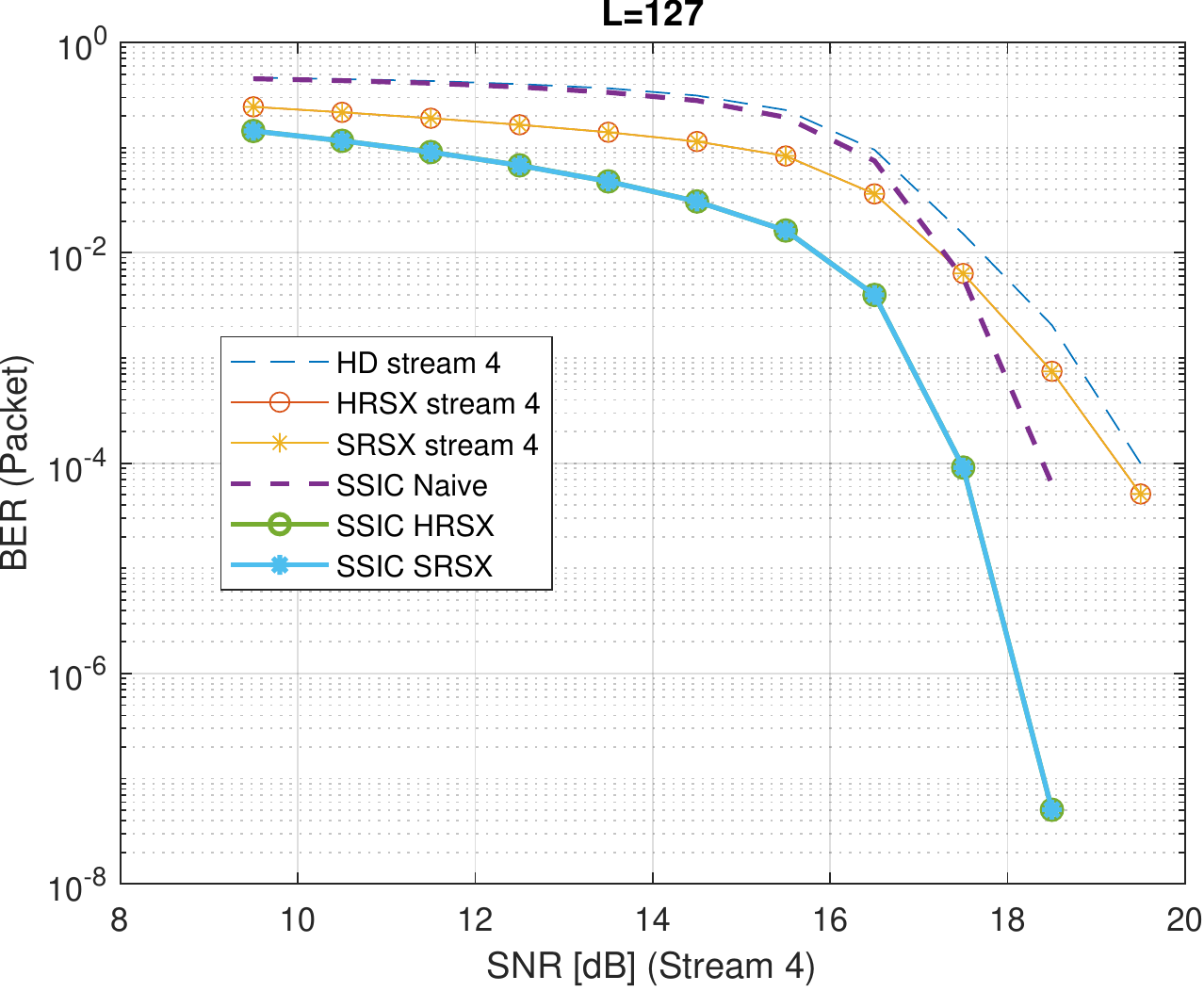}%
		\label{fig_10(b)}}
	
	\hfil
	
	\subfloat[]{\includegraphics[width=2.3in]{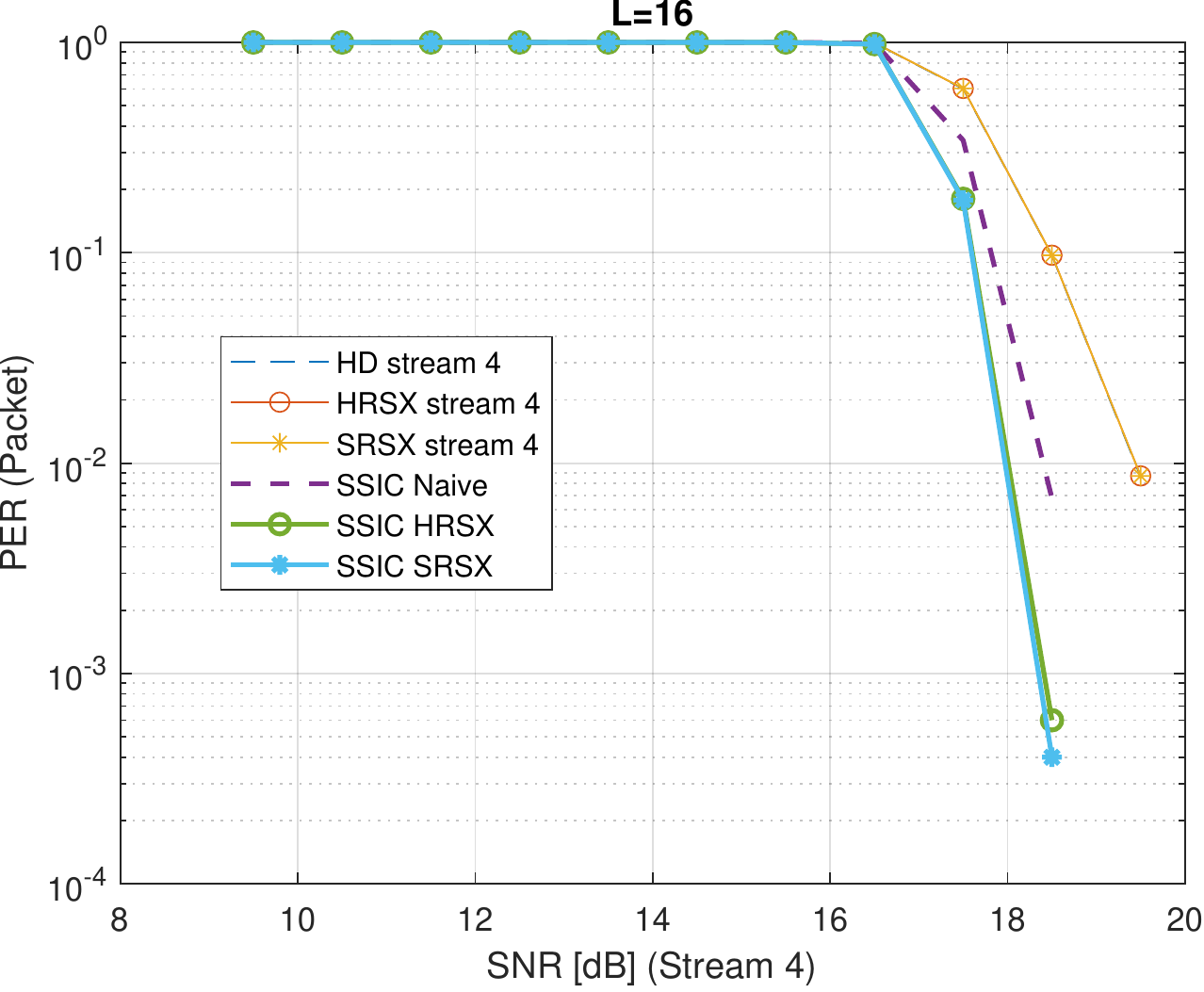}%
		\label{fig_10(c)}}
	
	\subfloat[]{\includegraphics[width=2.3in]{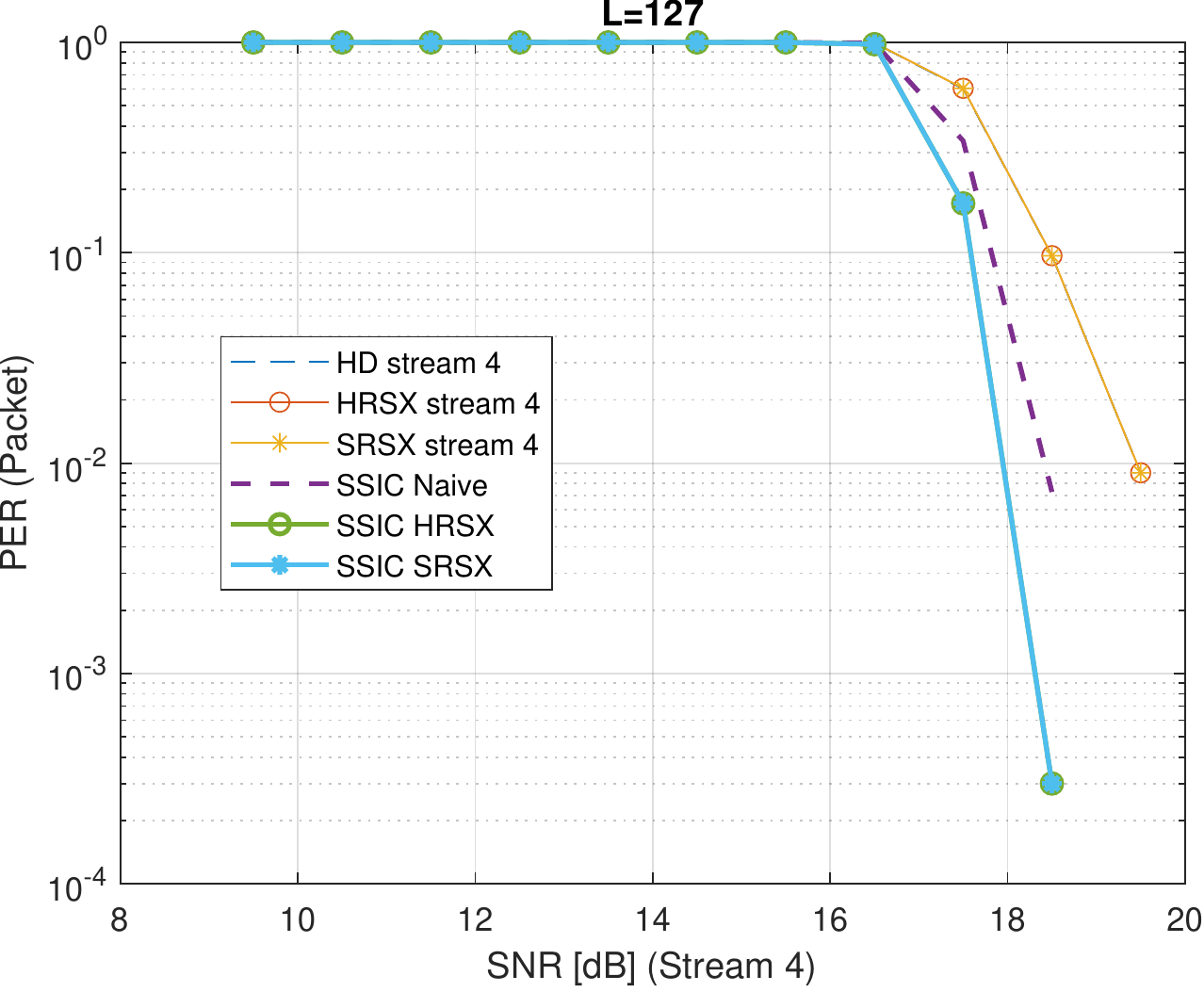}%
		\label{fig_10(d)}}
	
	\caption{BER and PER comparisons among SSIC with naïve SD, SSIC with HRSX, and SSIC with SRSX: (a) and (b) show the BER results when L=16 and L=127, respectively; (c) and (d) show the PER results when $L=16$ and $L=127$ respectively. In this experiment, four streams with different SNRs are used.}
	\label{fig_10}
\end{figure}

The simulation results are shown in Fig. \ref{fig_9} and Fig. \ref{fig_10}. Fig. \ref{fig_9} shows the BER of the single-stream case (results of stream 1, which is the worst case among the single-stream cases), whereas Fig. \ref{fig_10} shows the BER and PER performance for both the SSIC case and a single-stream case (results of stream $4$, which is the best case among the single-stream cases). We can see the following:
\begin{enumerate}[label=(\roman*)]
	\item Fig. \ref{fig_9(a)} shows that both HRSX and SRSX significantly improve the decoding of seed bits $\{ {r_0},{r_1},...,{r_6}\}$. With redundancy introduced $\{ {r_0},{r_1},...,{r_6}\}$, the larger $L$ the better the performance of HRSX and SRSX compared with HD. 
	\item Fig. \ref{fig_9(b)} shows that with a better estimation of $\{ {r_0},{r_1},...,{r_6}\}$, both HRSX and SRSX improve the BER of the packet in the single-stream case. However, HRSX and SRSX have almost the same BER performance. 
	\item Fig. \ref{fig_10} shows that SSIC with either HRSX or SRSX descrambling has better BER and PER performance than SSIC with naïve SD. In particular, when $L = 16$, SSIC with SRSX already has $1$ dB BER gain over SSIC with naïve SD, as shown in Fig. \ref{fig_10(a)}.
	\item Fig. \ref{fig_10} shows, in general, SSIC with SRSX outperforms SSIC with HRSX. With a larger $L$, the performance gap between HRSX and SRSX becomes smaller. In particular, HRSX approximates SRSX when $L=127$, as shown in Fig. \ref{fig_10(b)} and Fig. \ref{fig_10(d)}. 
\end{enumerate}

\begin{figure}[!htbp]
	\centering
	\includegraphics[width=3.5in]{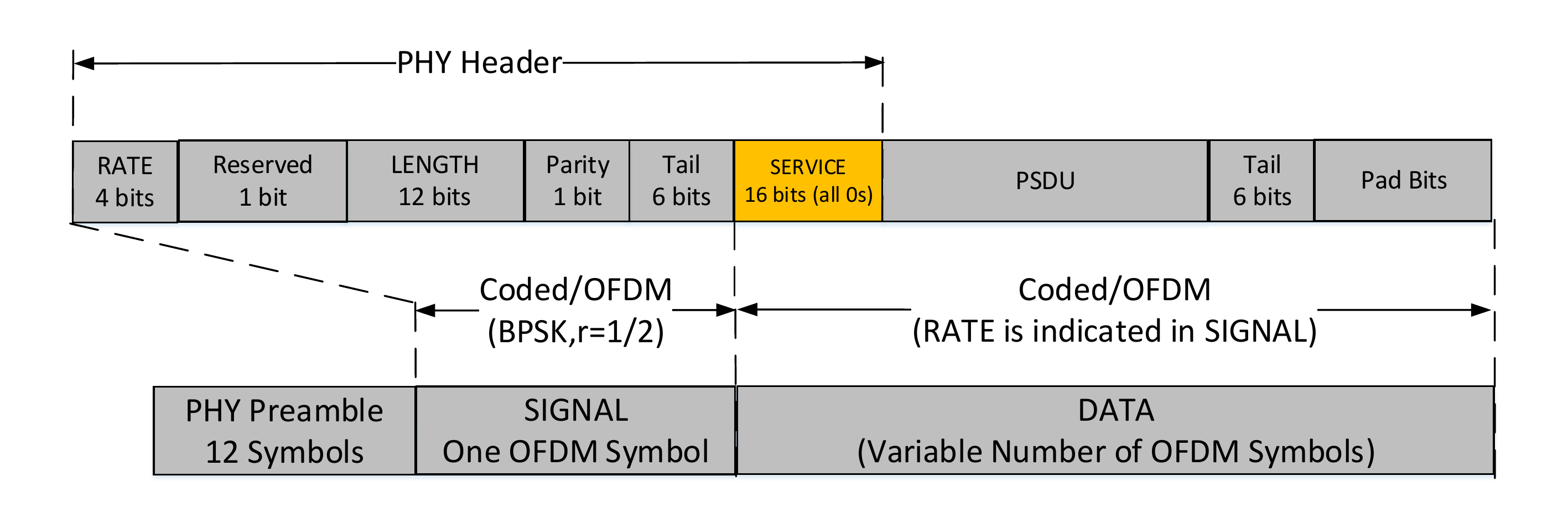}
	\caption{The PPDU format of legacy IEEE 802.11 in which 16 zeros bits already exist in the SERVICE field.}
	\label{fig_11}
\end{figure}

In short, with more than one stream’s soft information, SSIC with SD significantly improves communication reliability. What’s more, SRSX is preferred when SD is needed in SSIC since SRSX maintains good performance even when $L$ is small. For example, when $L=16$, as shown in Fig. \ref{fig_10}, SSIC with SRSX can already obtain good BER and PER performance and a further increase in $L$ is not necessary. Note that the legacy IEEE 802.11 PPDU already has 16 zeros bits in the SERVICE field as pilot bits, as shown in Fig. Fig. \ref{fig_11}. Overall, SRSX, $L=16$, is the best choice when deploying SSIC over Wi-Fi networks since it does not require any further pilot bits beyond those given by the standard. 

\

\noindent \textbf{\textit{Remark 3: Since the legacy 802.11 PPDU already has $16$ zeros bits in the SERVICE field as pilot bits, SRSX with $L=16$ is preferred when deploying SSIC in the Wi-Fi network because it maintains good performance when $L=16$ and does not require extra pilot bits and hence incurs zero additional overhead with respect to the standard.}}

\begin{figure}[!htbp]
	\centering
	\subfloat[]{\includegraphics[width=2.4in]{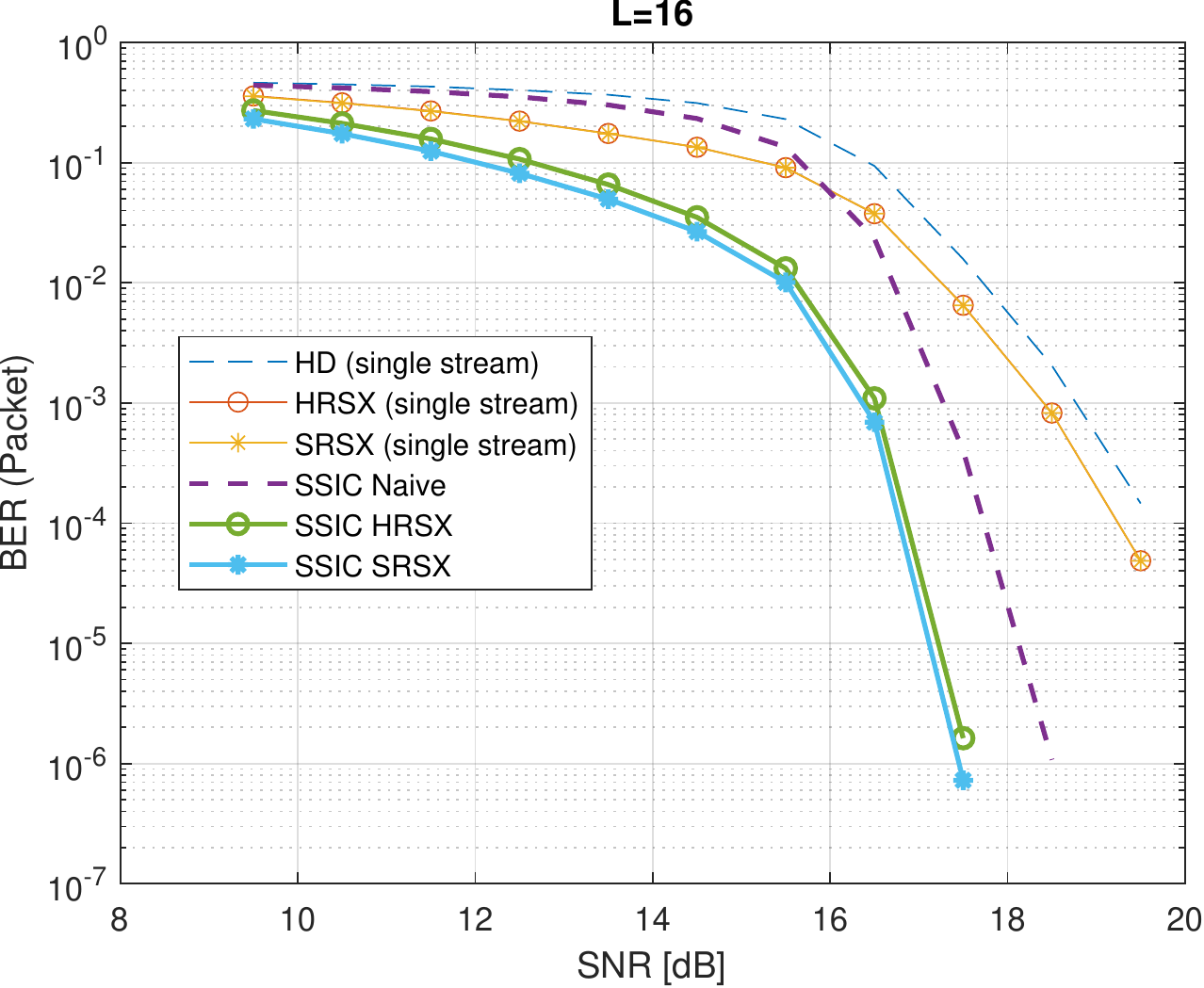}%
		\label{fig_12(a)}}
	
	\subfloat[]{\includegraphics[width=2.4in]{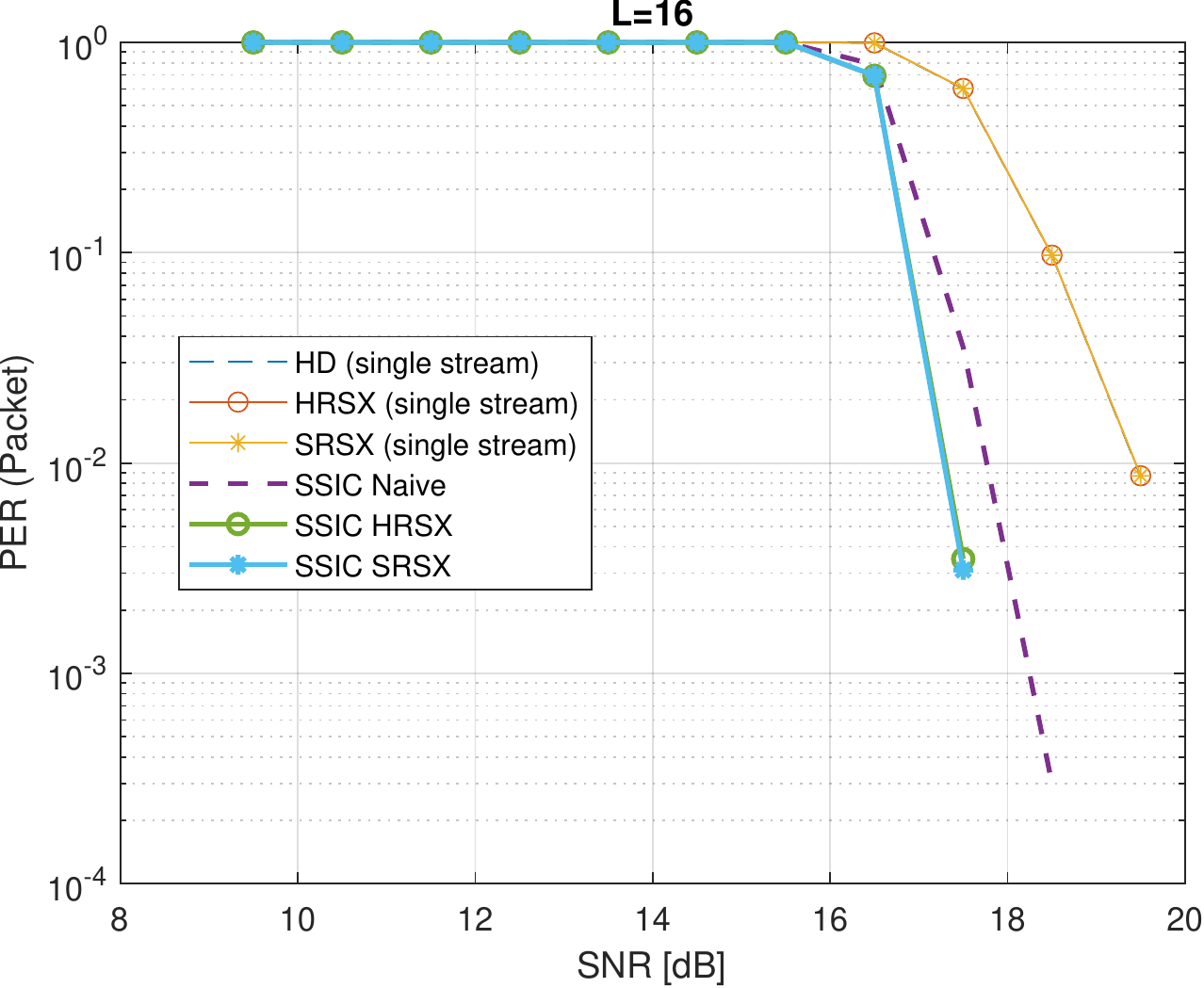}%
		\label{fig_12(b)}}
	\caption{BER comparisons among SSIC with naïve SD, SSIC with HRSX, and SSIC with SRSX when L=16: (a) shows the BER results; (b) shows the PER results. In this experiment, the four steams have the same SNR.}
	\label{fig_12}
\end{figure}

Let us look at another simulation setup with all four streams having the same SNR. We focus on the case with $L=16$. The BER and PER results are shown in Fig. \ref{fig_12(a)} and Fig. \ref{fig_12(b)}, respectively. We can see from these results that the improvement by SSIC is more obvious than in Fig. \ref{fig_10(a)} and Fig. \ref{fig_10(c)}, with more than $2$ dB BER and PER gains. In the next subsection, we investigate the performance of SSIC with SRSX in a real Wi-Fi network.

\subsection{SSIC network performance} \label{sec:SimulationExperiment:experiment}
\begin{figure}[!htbp]
	\centering
	\subfloat[]{\includegraphics[width=1.8in]{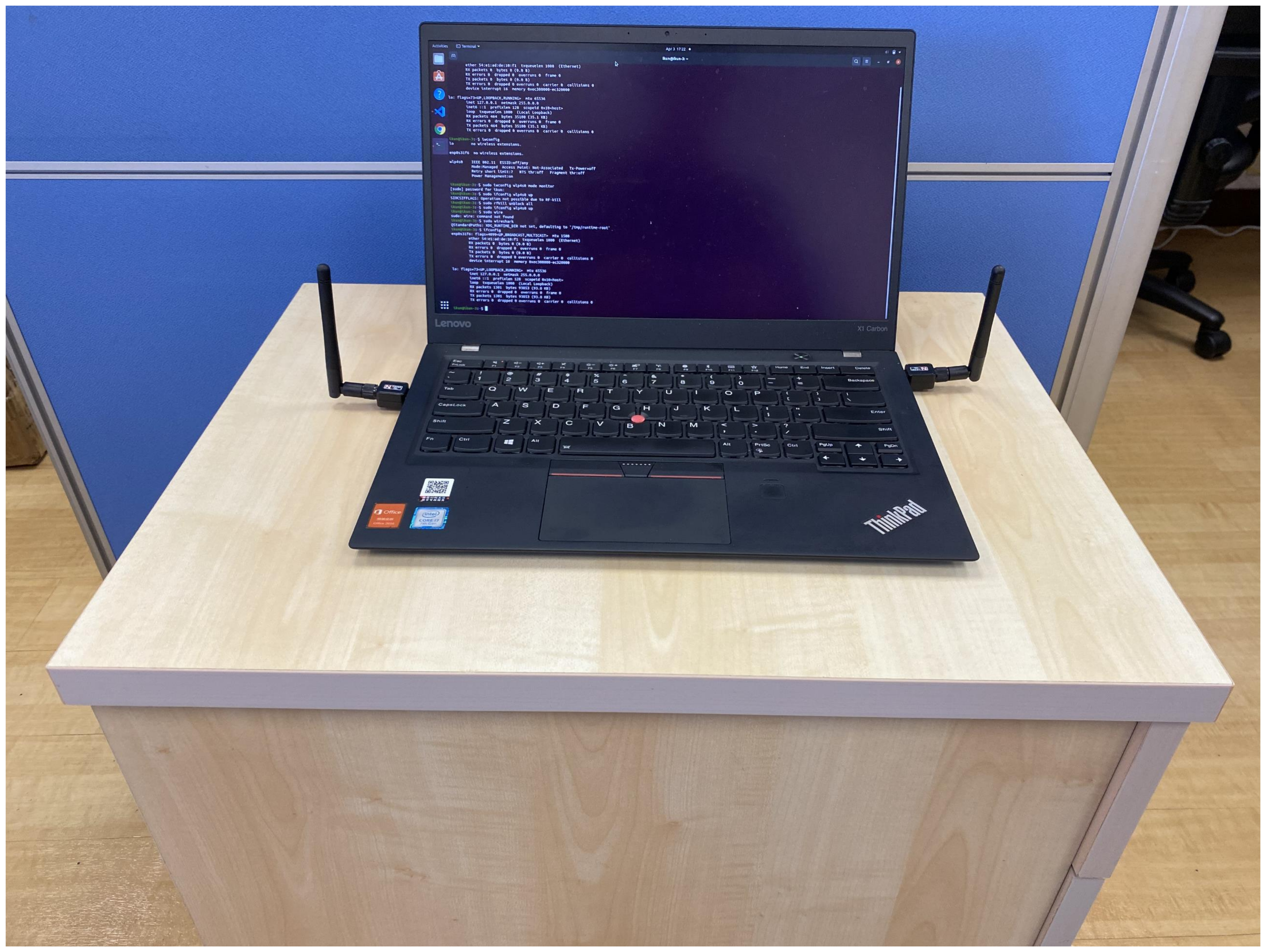}%
		\label{fig_13(a)}}
	\subfloat[]{\includegraphics[width=1.8in]{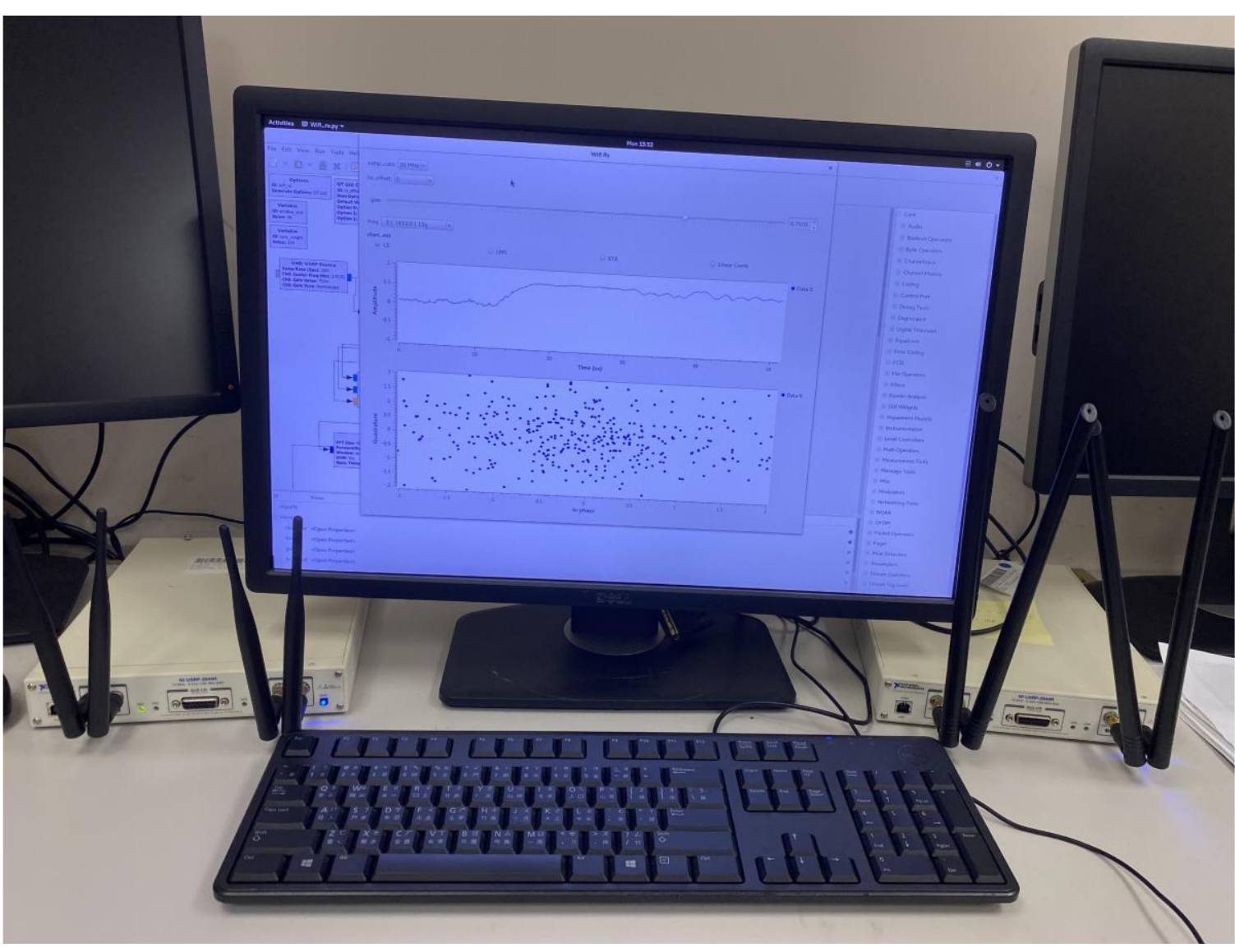}%
		\label{fig_13(b)}}
	
	\caption{Two PCs are set up to validate the functionality of the middleware: (a) PC A is an SSIC transmitter equipped with two Wi-Fi cards; (b) PC B is an SSIC receiver equipped with two USRPs}
	\label{fig_13}
\end{figure}

To validate the overall performance of SSIC in a real environment, we set up an SSIC network in our laboratory. Two general PCs are set up as shown in Fig. \ref{fig_13}: i) PC A, a laptop as shown in Fig. \ref{fig_13(a)}), serves as node A in Fig. \ref{fig_5}, and it is an SSIC transmitter with two commercial-used Wi-Fi USB sticks; ii) PC B, a PC as shown in Fig. \ref{fig_13(b)}, serves as node B in Fig. \ref{fig_5}, and it is an SSIC receiver equipped with two SDR devices – USRPs \cite{USRPX310}. Here each USRP attached to PC B simulates a NIC that is capable of outputting the output of $\tilde y_m^{}$ -- there is no such commercial NIC yet. The middleware runs on both nodes. Furthermore, the BCH (63,7) coding scheme is used to protect the VC header. Meanwhile, as suggested by Remark 3, the SRSX with $L=16$ is used for SD purposes.

We investigated the performance of SSIC networking over two noisy channels on the 2.4 GHz ISM band (one on WLAN channel 1, and another on WLAN channel 8). These two channels are noisy and interference-prone because a number of laptops, tablet computers, computer printers, and cellphones are also using the channels in our laboratory. The other settings are listed in Table \ref{tab:2}.

\begin{table}[!hbp]
	\caption{The detailed experiment settings}
	\label{tab:2}
	\begin{tabular}{|l|l|lll}
		\cline{1-2}
		ISM   band of stream 1                 & \begin{tabular}[c]{@{}l@{}}WLAN   channel 1 \\ (Center Freq: 2.412GHz)\end{tabular} &  &  &  \\ \cline{1-2}
		ISM   band of stream 2                 & \begin{tabular}[c]{@{}l@{}}WLAN   channel 8 \\ (Center Freq: 2.442GHz)\end{tabular} &  &  &  \\ \cline{1-2}
		PHY   type                             & 802.11a   OFDM                                                                      &  &  &  \\ \cline{1-2}
		Quantization of soft information       & 6-bit quantization \cite{volkhausen2012quantization}                    &  &  &  \\ \cline{1-2}
		Modulation   and Channel Coding scheme & 16-QAM, 1/2 Code Rate                                                &  &  &  \\ \cline{1-2}
		Coding scheme for VC header& BCH(63,7)                                              &  &  &  \\ \cline{1-2}
		Number of retransmission in 802.11 protocol & 0                                                     &  &  &  \\ \cline{1-2}
	\end{tabular}
\end{table}

Once set up, a UDP client running on PC A sends UDP packets to a UDP server running on PC B. In particular, each UDP packet is duplicated and encapsulated into two VC packets by the middleware on PC A (detailed in Section \ref{sec:SystemDesign:Networking}). These two VC packets are then separately transmitted by the two NICs. In the meantime, the two USRPs on PC B receive the packets sent from PC A over the two streams.

Two experiments with different device-placement profiles are performed. In each experiment, we compare SSIC with two single-stream cases – one in which only stream 1 is used, and the other in which only stream 2 is used –  and one multi-stream case with duplicate transmissions over both streams. In the duplicate-transmission case, referred to as DUP in the rest of the paper, packets are processed in a first-come-first-serve (FCFS) manner, and duplicated packets that arrive later, if any, are dropped (as in PRP).

\subsubsection{Experiment 1} \label{sec:SimulationExperiment:experiment:1}
\begin{figure}[!htbp]
	\centering
	\includegraphics[width=3.0in]{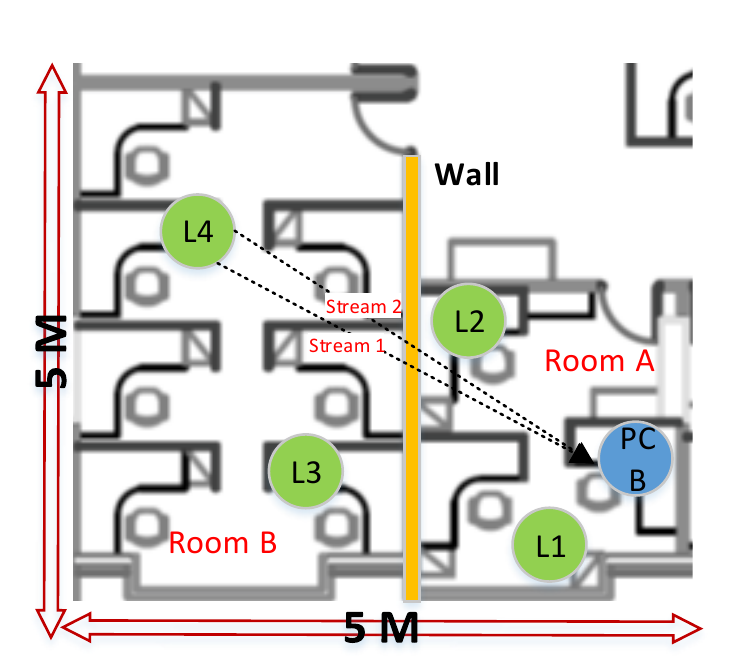}%
	
	\caption{In experiment 1, PC A was located in several different places (from location `L$1$' to location `L$4$') for the different rounds of the experiment, whereas PC B attached with USRPs was fixed at one place all the time.}
	\label{exp1:placement}
\end{figure}

\noindent In this experiment, PC A was located in four different places (location `L$1$', `L$2$', `L$3$' and `L$4$', as indicated in Fig. \ref{exp1:placement}) for the different rounds of the experiment, whereas PC B together with its attached two USRPs were always at the same place. This setup is as in Fig. \ref{fig_5(a)}, simulating the communication of two static devices.

We first investigated the VC-header decoding performance since the middleware relies on the VC header to identify a packet even when the packet does not pass the layer-2 CRC. To allow this investigation, the middleware on PC B collects the layer-2 CRC status and the VC-header CRC status for each packet. These statues are shown in Table \ref{tab:3}. We can see that the failure rate of the VC-header CRC is $10$x lower than that of layer-2 CRC. The VC header is well protected in the SSIC communication.

\begin{table*}[!htbp]
	\caption{VC CRC Failure rate VS Layer-2 CRC Failure rate in Experiment 1}
	\centering
	\label{tab:3}
	\begin{tabular}{|c|c|c|c|c|c|c|c|}
		\hline
		\multicolumn{1}{|l|}{Round No.} &
		\multicolumn{1}{l|}{\begin{tabular}[c]{@{}l@{}}PC A Location\end{tabular}} &
		\multicolumn{1}{l|}{\begin{tabular}[c]{@{}l@{}}Layer-2 CRC\\failure rate\\ (stream 1)\end{tabular}} &
		\multicolumn{1}{l|}{\begin{tabular}[c]{@{}l@{}}VC Header CRC\\failure rate\\ (stream 1)\end{tabular}} &
		\multicolumn{1}{l|}{\begin{tabular}[c]{@{}l@{}}Layer-2 CRC\\failure rate\\ (stream 2)\end{tabular}} &
		\multicolumn{1}{l|}{\begin{tabular}[c]{@{}l@{}}VC Header CRC\\failure rate\\ (stream 2)\end{tabular}}& 
		\multicolumn{1}{l|}{\begin{tabular}[c]{@{}l@{}}Distance Between\\PC A and PC B\end{tabular}}& 
		\multicolumn{1}{l|}{SNR}\\ \hline
		1 & L1 & 0.0017 & 2.04e-04 & 0.0013 & 3.4264e-04 & 1M LOS & 50dB \\ \hline
		2 & L2 & 0.0017 & 6.1782e-04 & 0.0093 & 5.8680e-04 & 2M LOS & 45dB   \\ \hline
		3 & L3 & 0.0107 & 0.0014 & 0.0097 & 6.9027e-04 & 3M NLOS & 35dB \\ \hline
		4 & L4 & 0.0095 & 0.0014 & 0.0206 & 0.0024 & 4M NLOS & 30dB\\ \hline
	\end{tabular}
\end{table*}

\begin{figure}[!htbp]
	\centering
	
	\subfloat[]{\includegraphics[width=2.6in]{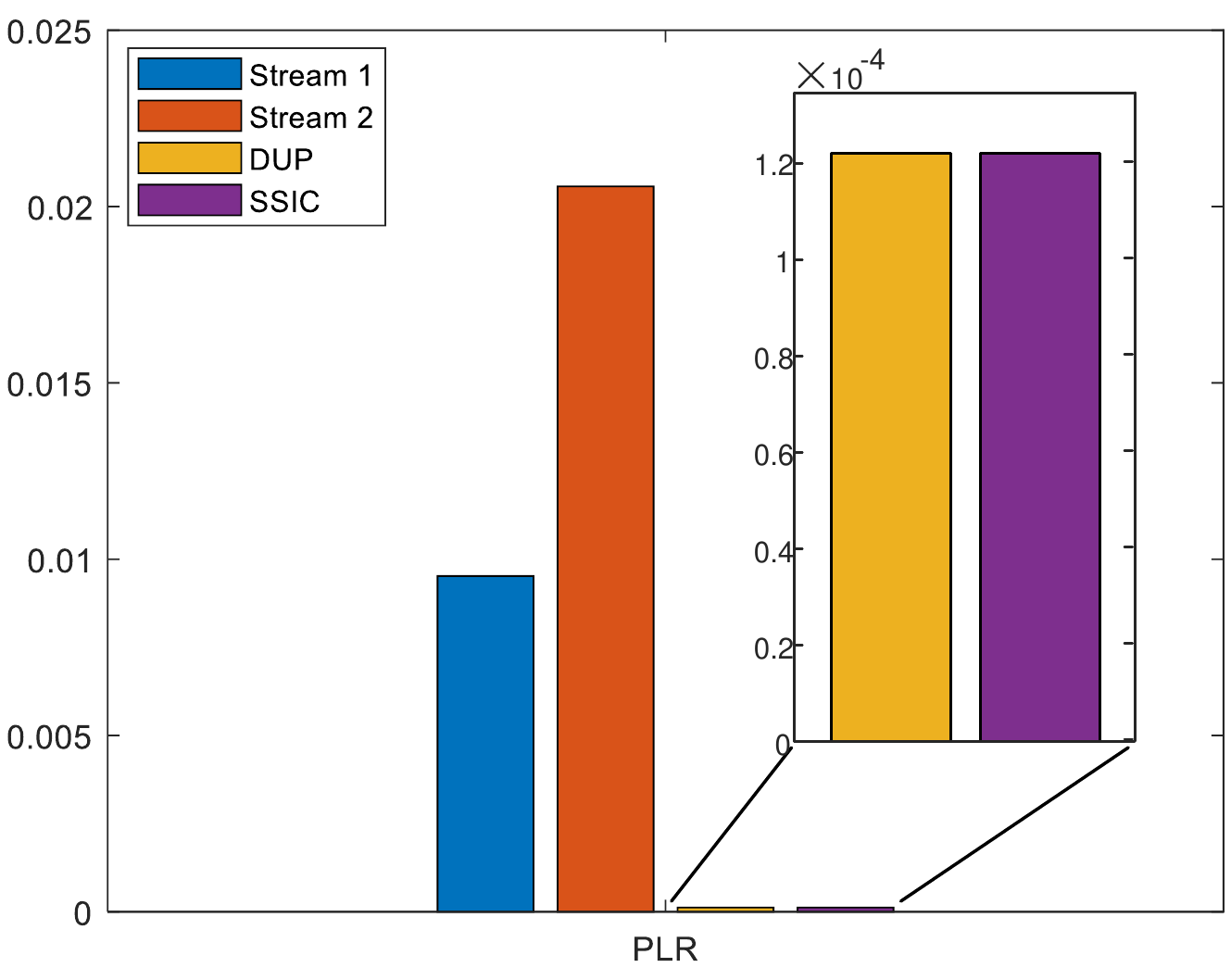}%
		\label{exp1:PLR}}
	
	\subfloat[]{\includegraphics[width=2.6in]{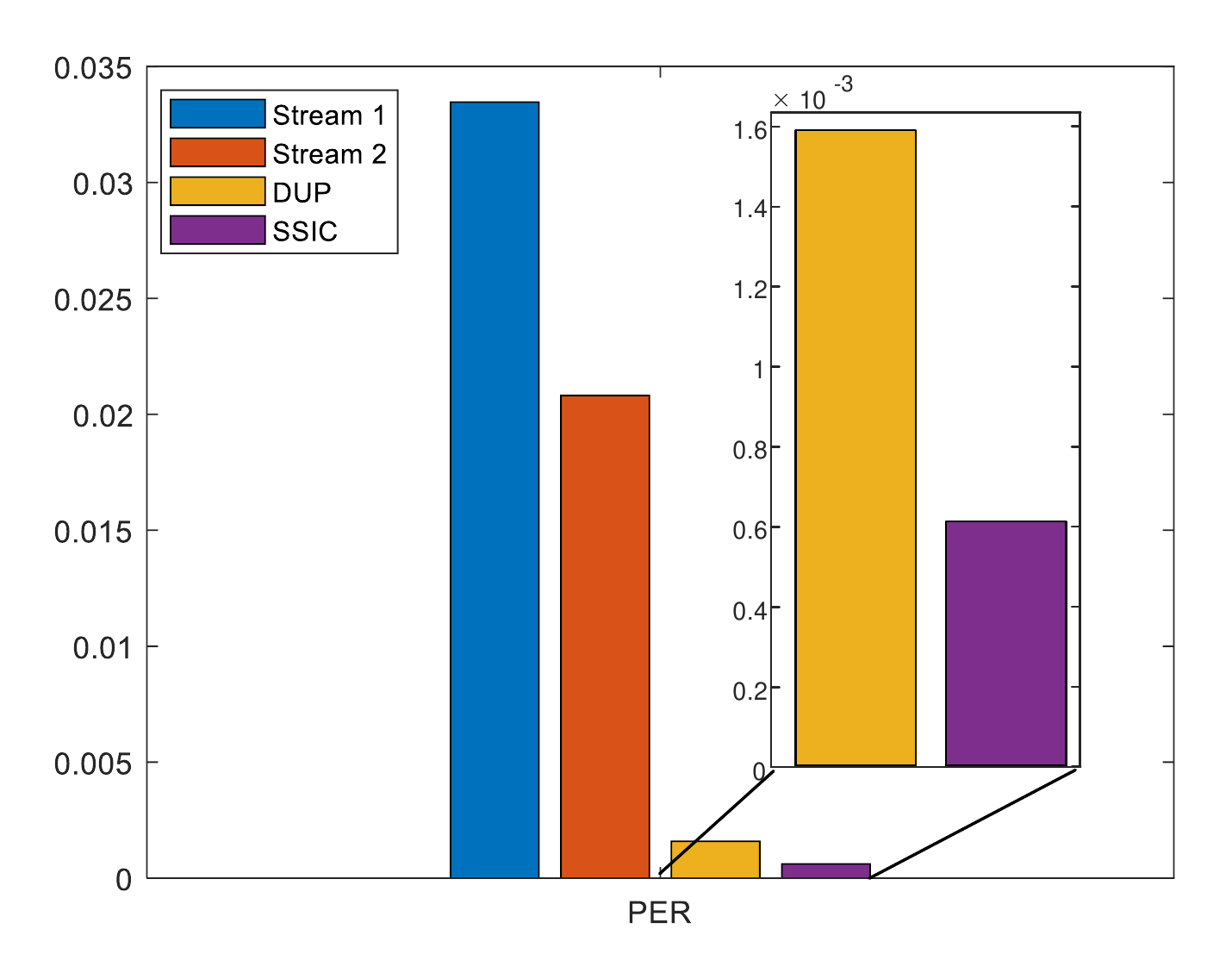}%
		\label{exp1:PER}}
	
	\subfloat[]{\includegraphics[width=2.6in]{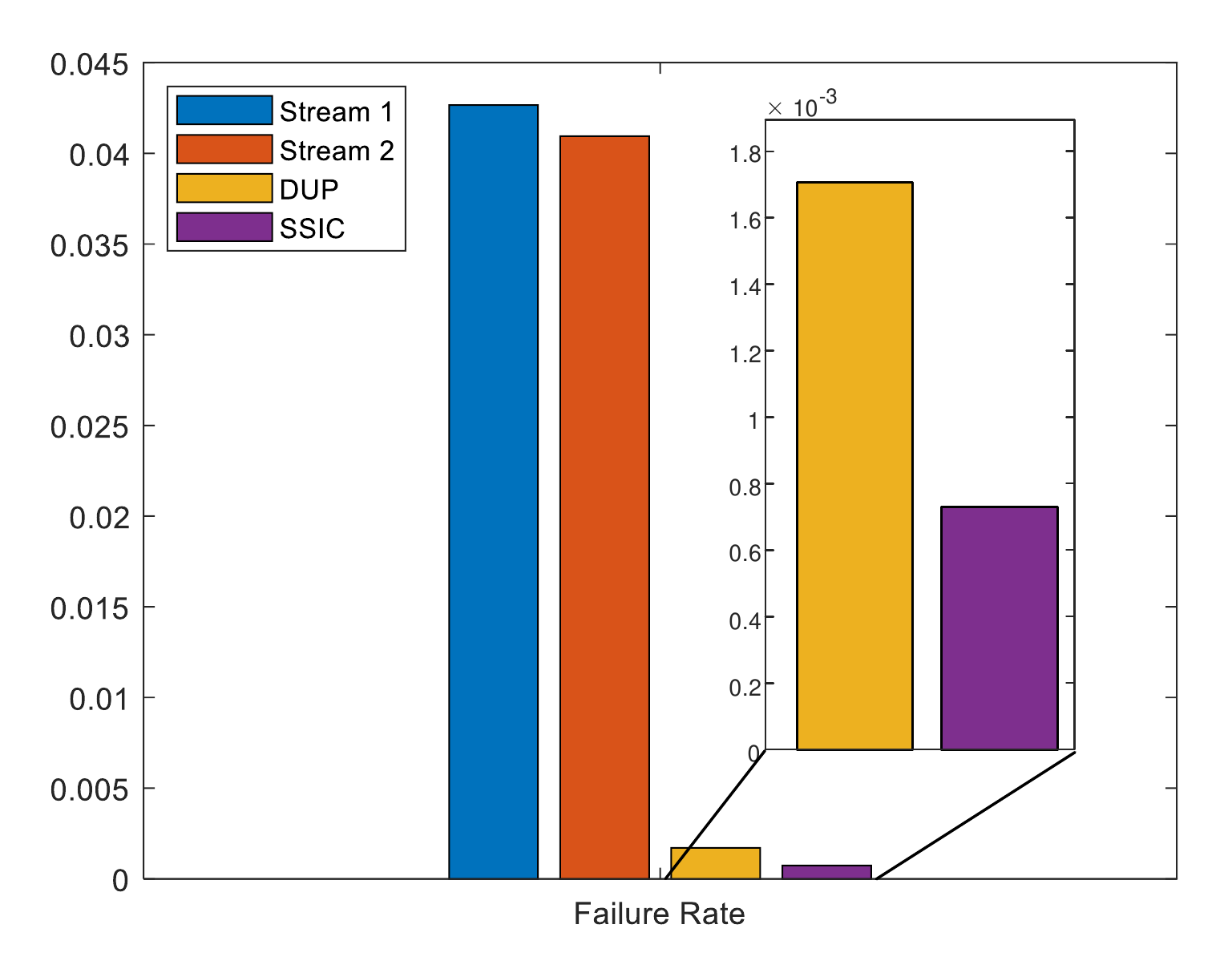}%
		\label{exp1:FR}}
	
	\caption{The PLR, PER, and FR of four cases (two single-stream cases, DUP and SSIC) are compared in experiment 1.}
	\label{exp1:PLRPERFR}
\end{figure}

We next investigated the packet-decoding performance of SSIC. Note that for Wi-Fi, only after the presence of a packet is detected will the packet decoding process kicks in. Hence, for a more thorough investigation, the middleware at PC B gathers statistics on packet loss rate (PLR) -- the packet miss-detection rate, packet error rate (PER) -- the probability of decoding failure for detected packets, and failure rate, $FR = 1 - (1 - PLR)(1 - PER)$, of the aforementioned four cases. Fig. \ref{exp1:PLRPERFR} shows the PLR, PER, and FR in round $4$ of the experiment, This is the case when the distance between PC A and B is the largest in experiment 1. The full experimental results for all four rounds can be found in Table \ref{tab:4}.

\begin{figure}[!htbp]
	\centering
	
	\subfloat[]{\includegraphics[width=2.6in]{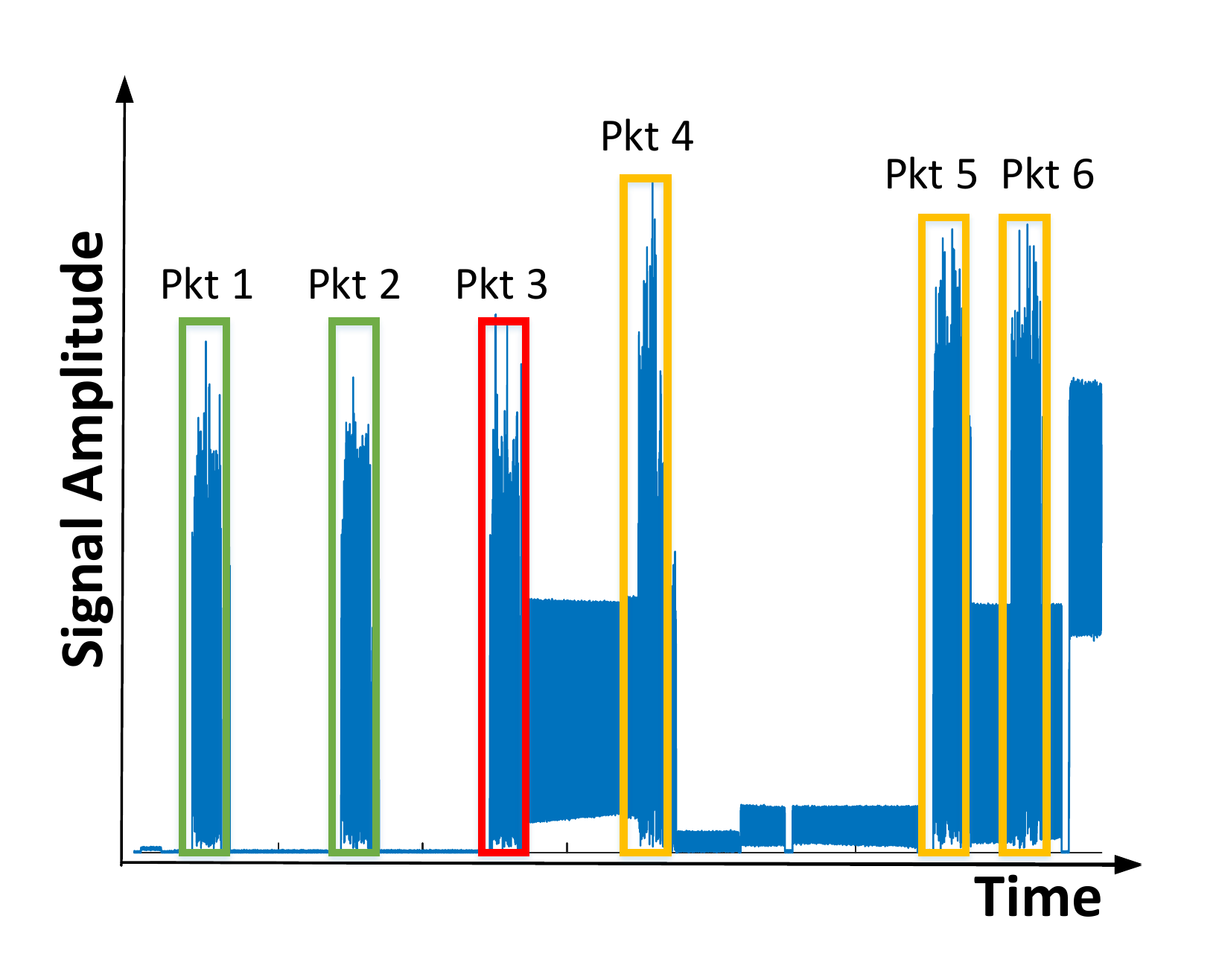}%
		\label{fig_17(a)}}
	
	\subfloat[]{\includegraphics[width=2.6in]{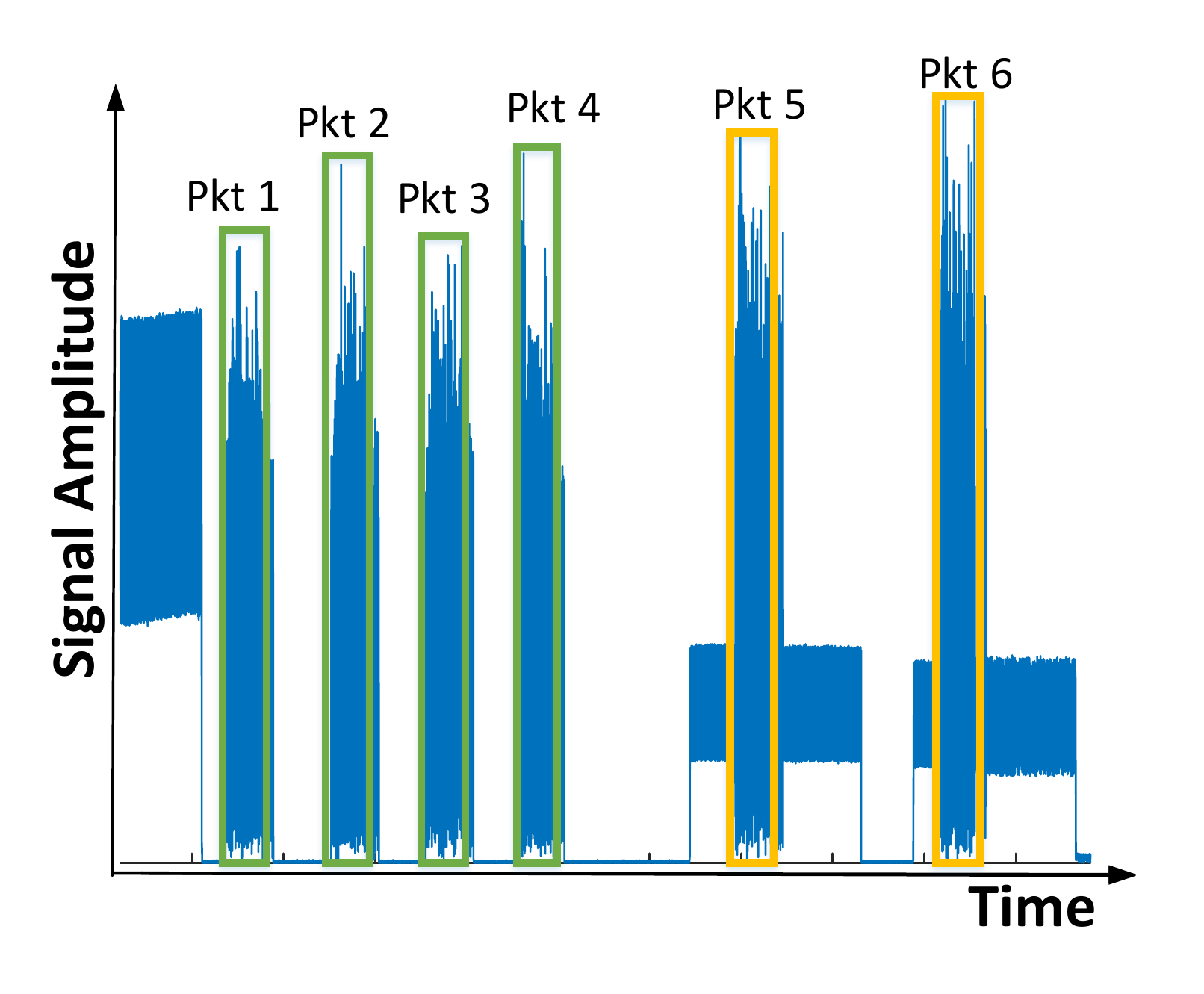}%
		\label{fig_17(b)}}
	
	\caption{Duplicates of seven packets are shown: (a) duplicate packets received from stream 1 (in particular, packets 3, 4, 5, and 6 experience the interference inflicted by other user’s packet); (b) duplicate packets received from stream 2 (in particular, packet 5 and 6 experience the interference inflicted by other user’s packet).}
	\label{fig_17}
\end{figure}

To better explain these results, we also plot the signal amplitudes of some received packets in round $4$ in Fig. \ref{fig_17}. In particular, Fig. \ref{fig_17(a)} and Fig. \ref{fig_17(b)} show some of the packets received by stream 1 and stream 2, respectively. For easy reference, we index the six shown packets by $1$ to $6$ in Fig. \ref{fig_17}. Any two packets with the same index have the same VCS and the same VN, and hence are used for SSIC. We explain the results in the following:

\begin{enumerate}[label=(\roman*)]
	\item In terms of PLR, SSIC and DUP share the same performance and significantly outperform the single-stream cases, as shown in Fig. \ref{exp1:PLR}. This result is intuitive, because for both, a packet is not detected (lost) only if it is neither detected in stream $1$ nor stream $2$. We explain this result through Fig. \ref{fig_17}. For example, although packet $3$ is totally distorted by interference and cannot be detected in stream $1$ (as circled in red in Fig. \ref{fig_17(a)}), it is well received in stream 2 (as circled in green in Fig. \ref{fig_17(b)}). With more than a stream, multi-stream transmission like SSIC and DUP have much lower PLR. 
	\item While DUP and SSIC have the same low PLR, in terms of PER among the detected packets, DUP does not have an improvement over the single-stream case. SSIC, thanks to the signal combination from the two streams, has a much lower PER, as shown in Fig \ref{exp1:PER}. We further explain this result through Fig. \ref{fig_17}. For example, packet $4$, $5$ and $6$ are not decoded successfully in either stream $1$ or stream $2$, as circled in yellow in Fig. \ref{fig_17(a)} and Fig. \ref{fig_17(b)}. However, by combining two copies of soft information, SSIC successfully decodes these packets. 
\end{enumerate}

Overall, SSIC networking has the lowest FR in the noisy environment and provides good packet delivery performance, as shown in Fig. \ref{exp1:FR}. Specifically, with a 3-meter non-line-of-sight (NLOS) distance between PC A and PC B, the FR of SSIC is $6$x lower than that of single-stream cases and is more than $2$x lower than that of DUP. What's more, as highlighted in Table \ref{tab:4}, SSIC provides 99.99\% reliable transmission when the distance between two end devices is below 3 meters.

\begin{table*}[]
	\centering
	\caption{PLR, PER, and FR of four cases in Experiment 1}
	\label{tab:4}
	\begin{tabular}{|c|c|c|c|c|c|c|c|c|c|c|c|c|}
		\hline
		& \multicolumn{4}{c|}{PLR}              & \multicolumn{4}{c|}{PER}              & \multicolumn{4}{c|}{FR}     \\ \hline
		Round No.& Stream 1 & Stream 2 & DUP    & SSIC   & Stream 1 & Stream 2 & DUP    & SSIC   & Stream 1 & Stream 2 & DUP    & SSIC   \\ \hline
		1   & 0.0017   & 0.0013   & 0      & 0      & 0.0034   & 0.0047   & 0      & 0      & 0.0051   & 0.0060   & 0      & \textbf{0}      \\ \hline
		2   & 0.0017   & 0.0093   & 0      & 0      & 0.0100   & 0.0090   & 0.0003 & 0 & 0.0117   & 0.0182   & 0.0003 & \textbf{0}\\ \hline
		3   & 0.0107   & 0.0097   & 0      & 0      & 0.0189   & 0.0143   & 0.0004 & 0.0001      & 0.0294   & 0.0238   & 0.0004 & \textbf{0.0001}       \\ \hline
		4   & 0.0095   & 0.0206   & 0.0001 & 0.0001 & 0.0335   & 0.0208   & 0.0016 & 0.0006 & 0.0427   & 0.0410   & 0.0017 & \textbf{0.0007} \\ \hline
	\end{tabular}
\end{table*}

\subsubsection{Experiment 2} \label{sec:SimulationExperiment:experiment:2}
\begin{figure}[!htbp]
	\centering
	\includegraphics[width=2.8in]{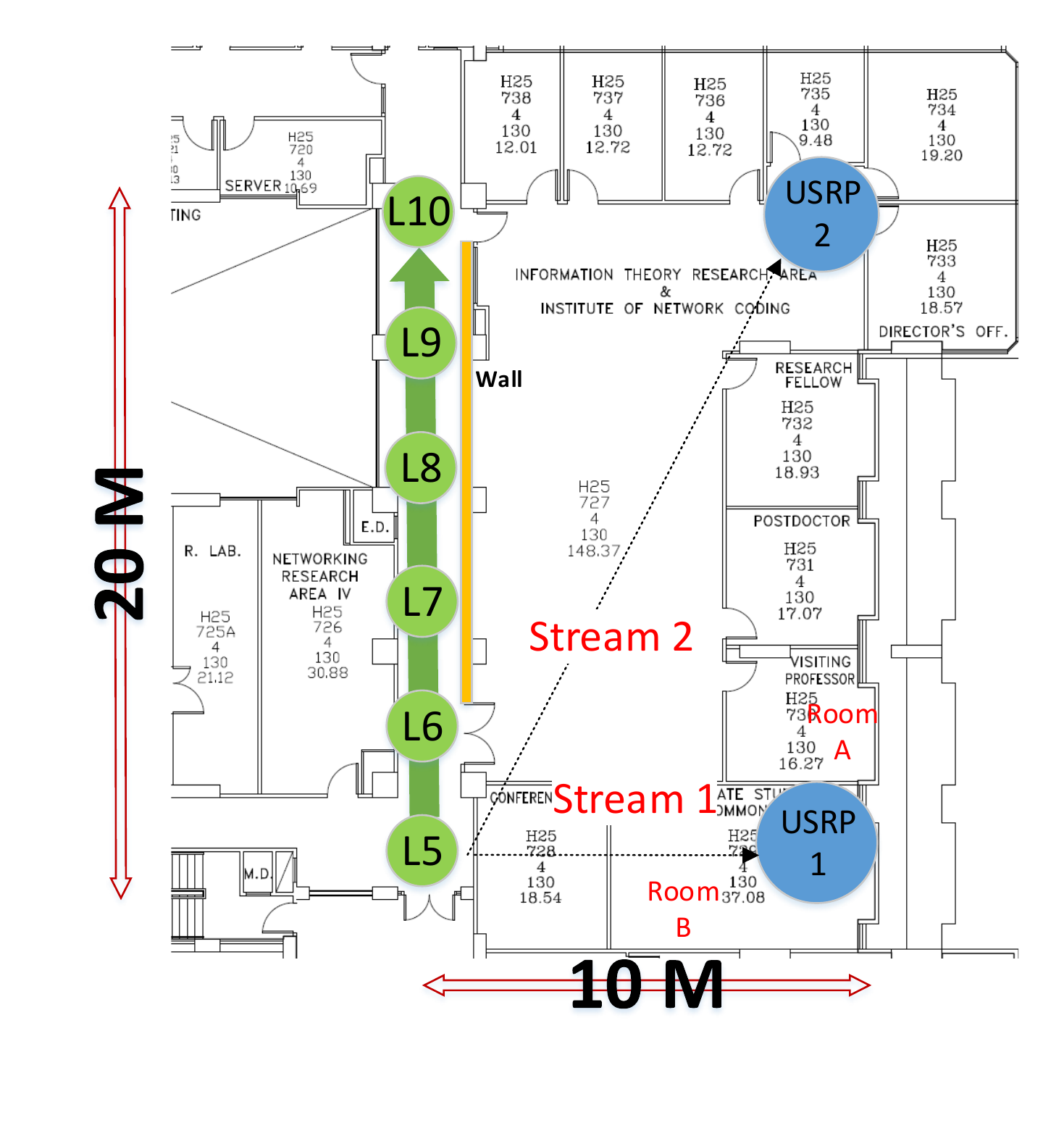}%
	\caption{In experiment 2, PC A was moving from location `L$5$' to location `L$10$', and two USRPs of PC B were located at different places.}
	\label{exp2:placement}
\end{figure}

In this experiment, PC A moved from location `L$5$' to location `L$10$', as illustrated in Fig. \ref{exp2:placement}. The two USRPs of PC B were located at different places. This setup is as in Fig. \ref{fig_5(b)}, simulating the case when one end device is moving while communicating with one or two access points (APs) simultaneously.

\begin{figure}[!htbp]
	\centering
	
	\subfloat[]{\includegraphics[width=2.5in]{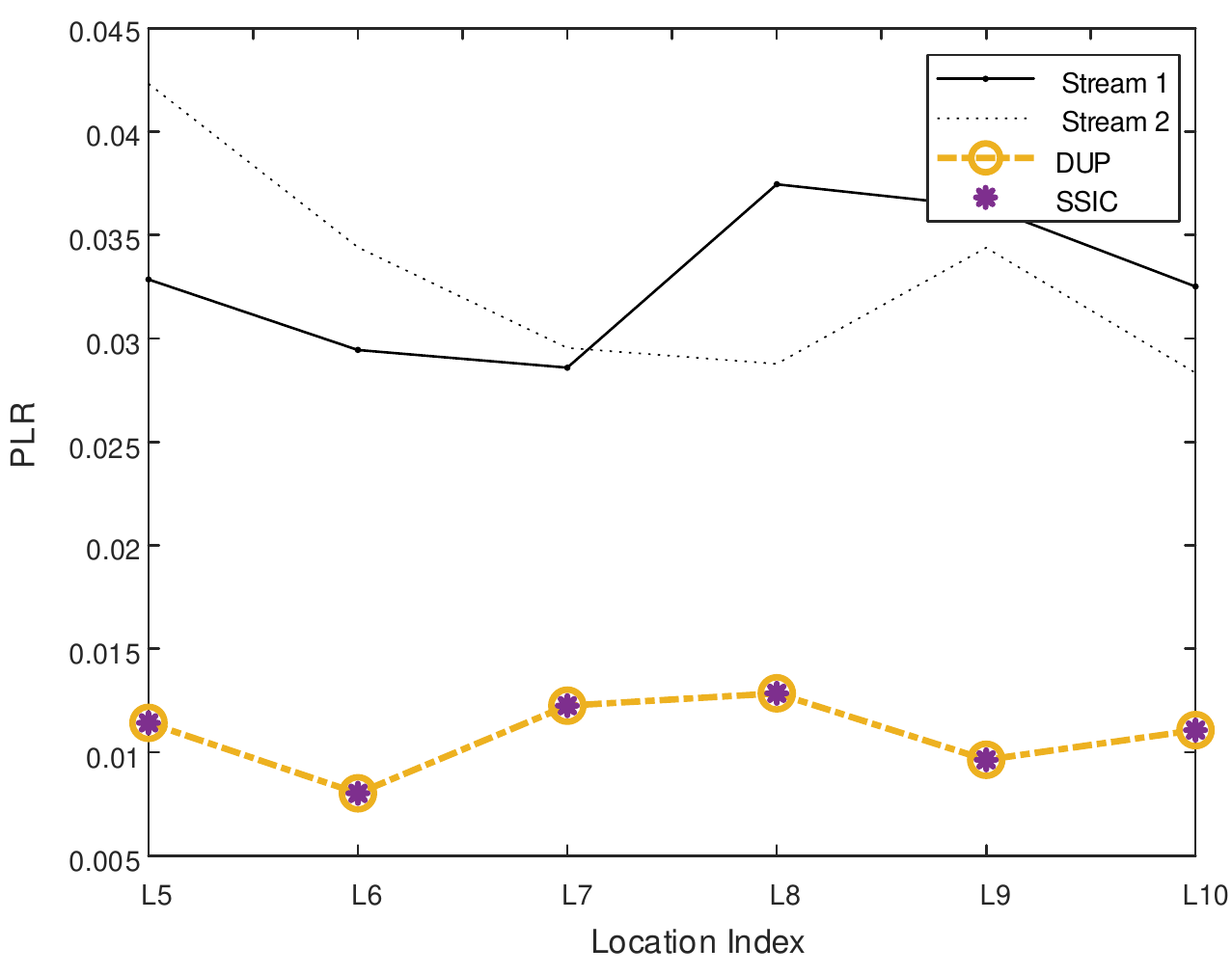}%
		\label{exp2-PLR}}
	
	\subfloat[]{\includegraphics[width=2.5in]{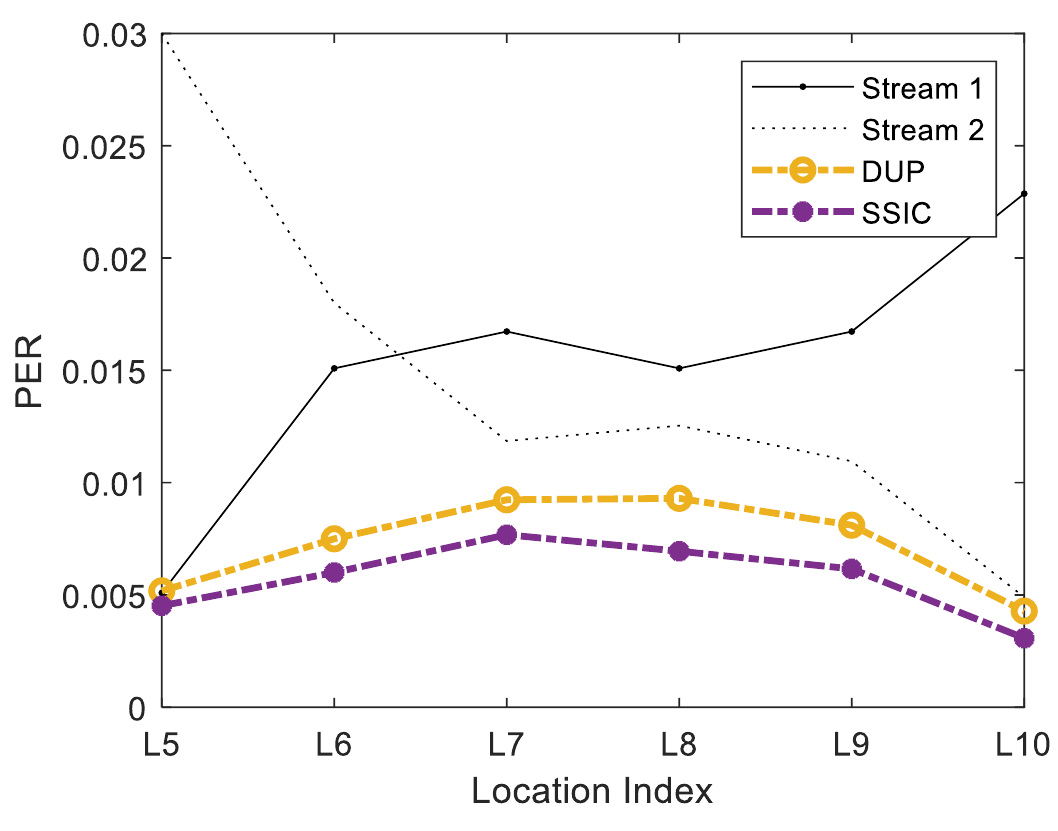}%
		\label{exp2-PER}}
	
	\subfloat[]{\includegraphics[width=2.5in]{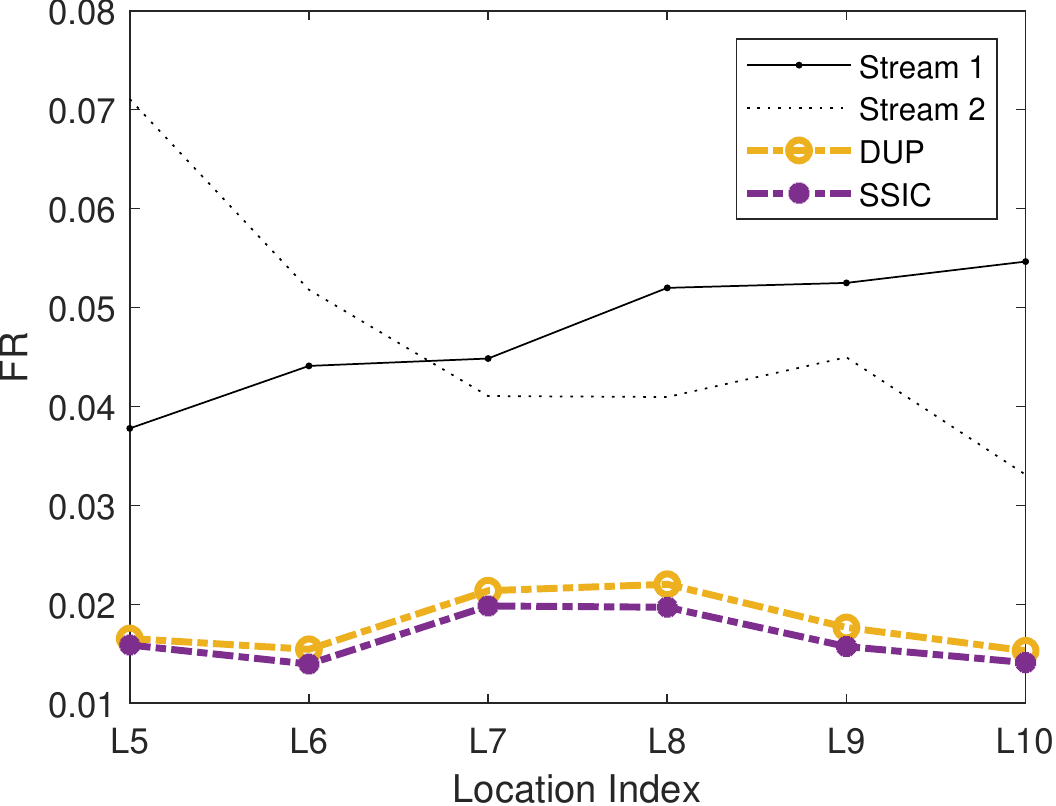}%
		\label{exp2-FR}}
	
	\caption{The PLR, PER, FR of four cases (two single-stream cases, DUP and SSIC) are compared in experiment 2.}
	\label{exp2:PLRPER}
\end{figure}

The experimental results, including PLR and PER, are shown in Fig. \ref{exp2:PLRPER} and described below::
	\begin{enumerate}[label=(\roman*)]
		\item In both single-stream cases, PLR fluctuates, and PER increases as the PC A moves away from its receiver. In particular, the variant of PER is large. For example, when only using stream 2, PER at location `L$5$' is around $6$x larger than that at location `L$10$', as shown in Fig. \ref{exp2-PLR}.
		\item SSIC and DUP reduce both PLR and PER, as shown in Fig. \ref{exp2-PLR}. For example, SSIC and DUP enable PC A to have a PLR of around $0.01$ from location `L$5$' to location `L$10$', lowering PLR up to $4$x than that of single-stream cases. Particularly, SSIC achieves the lowest PER among all the four cases, as shown in Fig. \ref{exp2-PER}.
	\end{enumerate}

Experiment 2 shows that SSIC networking stabilizes both PLR and PER even though the performance of single streams fluctuates due to the movement of the end device, achieving the lowest overall FR, as shown in Fig. \ref{exp2-FR}. Furthermore, Fig. \ref{exp2:PLRPER} indicates that if the PLR (rather than the PER) is the limit, then the improvement of SSIC over DUP is not much. In other words, if the packet detection process could be improved, then the advantage of SSIC over DUP would be more obvious. This observation offers us a direction for future work. For example, a cooperated packet-detection process over different streams can be designed so that once one stream has detected a packet, this stream would notify other streams to get ready for their receptions, reducing the PLR as much as possible. 


\section{Related Work} \label{sec:RelatedWork}
\cite{miu2005improving} proposed combining receptions from multiple transmitters to recover faulty packets. However, the method in \cite{miu2005improving} does not use soft information as we do. Specifically, \cite{miu2005improving} divides a packet into multiple blocks. When the transmitters receive conflicting blocks, \cite{miu2005improving} attempts to resolve the conflict by trying all block combinations and checking whether any resulting combinations produce a packet that passes the checksum test. \cite{miu2005improving} cannot help when conflicting blocks are all wrong. 

The method in \cite{jamieson2007ppr} uses the soft information to find incorrect chunks in a packet and retransmit only those chunks rather than the entire packet. However, this design is not compatible with existing NICs since legacy NICs are not allowed to transmit only a particular part of a packet. Moreover, \cite{jamieson2007ppr} does not have a multi-stream transmission mechanism.

The closest work to ours is \cite{woo2007beyond}, which proposed an architecture for WLAN to allow the physical layer to convey its soft source bits to the higher layers. A receiver can then combine the source bits from multiple transmitters to correct faulty bits in a corrupted packet. However, \cite{woo2007beyond} did not take care of the inherent scrambling functionality in the WLAN system. As stated in \textbf{\textit{Challenge 1}} in Section \ref{sec:introduction}, soft source bits in the WLAN system are scrambled randomly and need to be descrambled at the receiver side with their respective scrambling masks before they can be combined. Also, \cite{woo2007beyond} lacks a detailed networking design to resolve \textbf{\textit{Challenge 2}} in Section \ref{sec:introduction}. The implementation realized on USRPs in \cite{woo2007beyond} only processed the PHY-layer signal without validating whether the design is compatible with commercial NICs and TCP/IP networks. 

\section{Conclusion} \label{sec:Conclusion}
Existing duplicate-transmission standards, such as PRP in IEEE 802.11 or IEEE 802.1CB in the upcoming IEEE 802.11 be, lack a joint processing mechanism for duplicates, foregoing an effective way to boost reliability. This paper puts forth a soft-source-information-combining (SSIC) framework that combines the soft information of the duplicates to effect highly reliable communication. 
The SSIC framework contains two salient components:
\begin{enumerate}[label=(\roman*)]
	\item A soft descrambler (SD) that minimizes the bit-error rate (BER) and packet-error rate (PER) at the SSIC’s output –- SD yields the “soft” masking sequence required to descramble the soft information properly to minimize BER and PER of SSIC.
	\item An SSIC networking architecture readily deployable over today’s TCP/IP networks without specialized NICs –- The architecture allows multiple streams to be grouped and exposed as one single virtual link to the application for TCP/IP communication without the need to modify legacy TCP/IP applications.
\end{enumerate}

We set up an SSIC network testbed over Wi-Fi. Experimental results indicate that with two streams over two paths in a noisy and lossy wireless environment, the SSIC network can i) decrease the PER and packet loss rate (PLR) by more than fourfold compared with a single-stream network; ii) provide 99.99\% reliable packet delivery for short-range communication.

We believe that SSIC can be easily incorporated into Wi-Fi 7 to enhance the performance of its multi-link mode. Furthermore, SSIC can be also deployed in a heterogeneous networking set-up in which the different paths are established over wireless networks of different types (e.g., Wi-Fi and WiMAX).

\section*{Acknowledgments}
The authors acknowledges the invaluable assistance of Prof. He (Henry) Chen and thank Dr. Jiaxin Liang for his constructive suggestions on the prototype development.

\bibliographystyle{IEEEtran}
\bibliography{database}

\end{document}